\documentclass[11pt,draftclsnofoot,onecolumn]{IEEEtran}

\usepackage{amsmath}
\usepackage{amsfonts}
\usepackage{amssymb}
\usepackage{graphicx}
\usepackage{dsfont}
\usepackage{subfigure}
\usepackage{setspace}
\usepackage{cite}

\DeclareMathOperator*{\argmin}{arg\,min}

\newcommand{\tr}{\mathop{\mathrm{tr}}\nolimits}
\newcommand{\vect}{\mathop{\mathrm{vec}}\nolimits}
\newcommand{\rank}{\mathop{\mathrm{rank}}\nolimits}

\newtheorem{Remark}{Remark}

\def\cF{\mathcal{F}}

\def\cI{\mathcal{I}}

\def\cK{\mathcal{K}}

\def\cM{\mathcal{M}}
\def\cN{\mathcal{N}}

\def\bcG{\boldsymbol{\mathcal{G}}}
\def\bcH{\boldsymbol{\mathcal{H}}}

\def\bcT{\boldsymbol{\mathcal{T}}}
\def\bcU{\boldsymbol{\mathcal{U}}}

\def\bcW{\boldsymbol{\mathcal{W}}}

\def\ba{{\mathbf{a}}}

\def\bc{{\mathbf{c}}}

\def\bff{{\mathbf{f}}}

\def\bh{{\mathbf{h}}}

\def\bn{{\mathbf{n}}}

\def\bs{{\mathbf{s}}}

\def\bu{{\mathbf{u}}}

\def\bx{{\mathbf{x}}}
\def\by{{\mathbf{y}}}

\def\b0{{\mathbf{0}}}

\def\bA{{\mathbf{A}}}
\def\bB{{\mathbf{B}}}
\def\bC{{\mathbf{C}}}
\def\bD{{\mathbf{D}}}
\def\bE{{\mathbf{E}}}
\def\bF{{\mathbf{F}}}
\def\bG{{\mathbf{G}}}
\def\bH{{\mathbf{H}}}
\def\bI{{\mathbf{I}}}

\def\bR{{\mathbf{R}}}

\def\bU{{\mathbf{U}}}
\def\bV{{\mathbf{V}}}
\def\bW{{\mathbf{W}}}
\def\bX{{\mathbf{X}}}
\def\bY{{\mathbf{Y}}}
\def\bZ{{\mathbf{Z}}}

\def\bbC{{\mathbb{C}}}

\def\bbE{{\mathbb{E}}}

\def\rF{\mathrm{F}}

\def\rR{\mathrm{R}}

\def\rT{\mathrm{T}}
\def\rU{\mathrm{U}}

\def\rX{\mathrm{X}}

\def\rm{\mathrm{m}}

\def\rx{\mathrm{x}}

\def\rmax{\mathrm{max}}

\def\rMMSE{\mathrm{MMSE}}

\def\rsum{\mathrm{sum}}

\def\ropt{\mathrm{opt}}

\def\rPC{\mathrm{PC}}
\def\rnoPC{\mathrm{noPC}}

\def\cOP{\mathcal{OP}}

\def\cTL{\mathcal{TL}}
\def\cWMSE{\mathcal{WMSE}}

\begin{document}

\title{Cooperative Algorithms for MIMO Amplify-and-Forward Relay Networks}
\author{
Kien~T.~Truong,~\IEEEmembership{Student Member,~IEEE,} Philippe~Sartori, and Robert~W.~Heath,~Jr.*,~\IEEEmembership{Fellow,~IEEE}
\thanks{The authors are with the Department of Electrical and Computer Engineering, WNCG, 2501 Speedway Stop C0806, The University of Texas at Austin, Austin, TX, 78712-1687 (email: kientruong@utexas.edu and rheath@ece.utexas.edu, phone: (512) 686 8225, fax: (512) 471 6512). P. Sartori is with Huawei Technologies, Inc. (email: philippe.sartori@huawei.com)}
\thanks{This work was supported by a gift from Huawei Technologies, Inc.}}
\markboth{Submitted to IEEE Transactions on Signal Processing}
{}\maketitle

\begin{abstract}
Interference alignment is a signaling technique that provides high multiplexing gain in the interference channel. It can be extended to multi-hop interference channels, where relays aid transmission between sources and destinations. In addition to coverage extension and capacity enhancement, relays increase the multiplexing gain in the interference channel. In this paper, three cooperative algorithms are proposed for a multiple-antenna amplify-and-forward (AF) relay interference channel. The algorithms design the transmitters and relays so that interference at the receivers can be aligned and canceled. The first algorithm minimizes the sum power of enhanced noise from the relays and interference at the receivers. The second and third algorithms rely on a connection between mean square error and mutual information to solve the end-to-end sum-rate maximization problem with either equality or inequality power constraints via matrix-weighted sum mean square error minimization. Since we can find a globally optimal solution in each iteration, the resulting iterative algorithms are convergent. Simulations show that the proposed algorithms achieve higher end-to-end sum-rates and multiplexing gains that existing strategies for AF relays, decode-and-forward relays, and direct transmission. The first algorithm outperforms the other algorithms at high signal-to-noise ratio (SNR) but performs worse than them at low SNR. Thanks to power control, the third algorithm outperforms the second algorithm at the cost of additional overhead.
\end{abstract}


\section{Introduction}\label{sec:Intro}
Relay interference channels model networks where a stage of intermediate nodes, called relays, help multiple transmitters communicate with their receivers using shared radio resources~\cite{SimeoneEtAl2007:Allerton, ThejaswiEtAl2008:Allerton, CaoChen2009:Allerton, TorabiFrigon2008:Globecom}. Upcoming cellular standards are considering relay communication for coverage extension and capacity enhancement~\cite{3GPPLTEAdvanced:STANDARD,IEEE80216m:STANDARD}. Prior work, however, shows that single-antenna relays do not work well in the presence of co-channel interference~\cite{ZhuZheng2008:TMC,PetersEtAl2009:JWCN}. In this paper, we consider multiple-antenna relay systems to take the advantage of the interference management capability of multiple-input multiple-output (MIMO) communication. Many interference management strategies have been proposed for the MIMO single-hop interference channel~\cite{CadambeJafar2008:TIT, GouJafar2010:TIT, YetisEtAl2010:TSP, RazaviyaynEtAl2011:TIT, Jafar2011:TUTORIAL}. 
Although these single-hop results can be applied separately for the transmitter-relay hop and for the relay-receiver hop, even higher sum-rates can be achieved if the relays are configured jointly~\cite{GhozlanEtAl2009:ISIT, JeonEtAl2011:TIT}. Obtaining the most from relay interference channels requires advanced interference management strategies that jointly configure the transmitters, relays, and receivers.

A general challenge to designing algorithms for the interference channel is that the sum capacity is unknown. The multiplexing gain of a network is a first-order approximation of its sum-capacity at high signal-to-noise ratio (SNR)~\cite{Host-MadsenNosratinia2005:ISIT}. Interference alignment is a multiplexing gain maximizing signaling technique for the single-hop interference channel~\cite{Jafar2011:TUTORIAL}, achieving the maximum number of degrees of freedom. The idea is to arrange the transmitted signals such that interference is constrained within only a portion of the signal space observed by each receiver, leaving the remaining portion for interference-free detection of the desired signal~\cite{CadambeJafar2008:TIT}. The maximum multiplexing gains achievable through interference alignment, however, depend on the characteristics of the interference channels. For a symmetric MIMO interference channel with constant channel coefficients, the maximum multiplexing gain is upper-bounded by the total number of antennas at a transmitter-receiver pair regardless the number of pairs~\cite{YetisEtAl2010:TSP, RazaviyaynEtAl2011:TIT}. Note that the bound is tight in certain cases and corresponds to the total available spatial dimensions of a pair. Increasing the number of spatial dimensions in the network, using for example relays, is one way to improve the maximum achievable multiplexing gain.

Relays can be classified based on their signal processing operation, among which the most popular are decode-and-forward (DF - the relays decode the received signals then re-encode before retransmitting) and amplify-and-forward (AF - the relays apply linear signal processing to the signal before forwarding). Without decoding the received signals, AF relays need no knowledge of the codebooks used by the transmitters and likely have lower baseband complexity and fast signal processing. In addition, transparent to the modulation and coding of the signals, AF relays are more suitable for applications in heterogeneous networks comprising many nodes of different complexity or even standards~\cite{BergerEtAl2009:CM}. In this paper, we focus on a half-duplex MIMO AF relay interference channel. Since half-duplex relays cannot transmit and receive at the same time, they are more practical than full-duplex relays.

Several interference management strategies designed specifically for the one-way AF relay interference channel have been proposed~\cite{JeonEtAl2011:TIT, CadambeJafar2009b:TIT,  GomadamEtAl2011:TIT, AbeEtAl2006:EURASIP, MorgenshternBolcskei2007:TIT, RankovWittenben2007:JSAC, NouraniEtAl2010:ISIT, ChenCheng2010:Globecom, ChaabanSezgin2010:CISS, GouEtAl2010:Arxiv, LeeJafar2011:ArXiv, ChaliseVandendorpe2010:TSP, NingEtAl2010:ISIT,LiangFeng2012:TVT,ZapponeJorswieck2012:TWC,ShiEtAl2012:TVT}. Although relays cannot improve the multiplexing gains of the single-antenna fully-connected interference channel with time-varying or frequency-selective channel coefficients~\cite{CadambeJafar2009b:TIT}, they are beneficial for reducing the number of independent channel extensions needed to align interference at the receivers~\cite{GomadamEtAl2011:TIT}. Prior work often considers networks operating in special circumstances. It is assumed in \cite{JeonEtAl2011:TIT, AbeEtAl2006:EURASIP, MorgenshternBolcskei2007:TIT, RankovWittenben2007:JSAC, NouraniEtAl2010:ISIT, ChenCheng2010:Globecom} that there are enough antennas at the relays to cancel all interference on the reception and then to nullify all interference on the retransmission, allowing multiplexing gains to scale linearly with the number of users. Other prior work considers only small networks with up to three pairs to derive some kind of closed-form strategies~\cite{ChaabanSezgin2010:CISS, GouEtAl2010:Arxiv, LeeJafar2011:ArXiv}. Prior work in~\cite{ChaliseVandendorpe2010:TSP} considers design problems with different objective functions including sum power minimization and minimum SINR maximization. In addition, the algorithms in~\cite{ChaliseVandendorpe2010:TSP} are applicable only for single-antenna receivers. Prior work in~\cite{LiangFeng2012:TVT,ZapponeJorswieck2012:TWC,ShiEtAl2012:TVT} develops noncooperative resource allocation strategies for AF relay networks while our work focuses on centralized algorithms for cooperative resource allocation. In this paper, we consider a general setting in the sense that we assume no special constraints on the number of wireless nodes or the number of antennas at a node. The closest AF relay model to ours is considered in~\cite{NingEtAl2010:ISIT}, which is only for single-antenna transmitters and receivers. Further, they assume no crosslinks from relays to receivers, resulting in an oversimplified design problem.

Sum-rate maximization problems are nonconvex and NP-hard, i.e., their global optima cannot be found in a polynomial time. It is even challenging to find their good local optima corresponding to interference aligned solutions using gradient-based algorithms from arbitrary initializations because those solutions have very narrow regions of attraction~\cite{SchmidtEtAl2009:Asilomar}. Thus, we propose to formulate three new design problems that have exactly the same constraints with sum-rate maximization problems but with better-behaved objective functions to find high-quality solutions in terms of sum-rate maximization. Based on the observation that interference alignment solutions make total leakage power go to zero~\cite{PetersHeath2008:ICASSP, PetersHeath2011:TVT, GomadamEtAl2011:TIT}, we formulate a problem that aims at minimizing the sum power of  interference and enhanced noise from the relays. Based on a relationship between achievable rates and mean squared error (MSE), we formulate two matrix-weighted sum-MSE minimization problems, either without power control or with power control. The key is that they have the same stationary points as their corresponding end-to-end sum-rate maximization problems. Although the newly formulated optimization problems are still nonconvex and NP-hard, they may be easier to solve. Next, we propose to adopt an alternating minimization approach~\cite{Bertsekas1999:BOOK,BezdekHathaway2003:NPSC} to develop iterative algorithms for solving the newly formulated problems. In each iteration, all but one variable is fixed and we focus on designing the remaining variable by solving a single-variable optimization problem obtained from the corresponding original multi-variable problems. Since we are able to find a global optimum for the single-variable optimization problem in each iteration, the proposed algorithms are guaranteed to converge. Note that the power constraints at the relays depend on both the transmit precoders at the transmitters and the processing matrices at the relays, adding more constraints to the design problems. Thus, it is not straightforward to extend the methods used for the single-hop design problems to solve the two-hop design problems. Our initial results in this paper were reported in~\cite{TruongHeath2011:Asilomar}. Compared with~\cite{TruongHeath2011:Asilomar}, this paper presents three different algorithms, has more discussion of convergence and provides simulations that emphasize the achievable end-to-end sum-rates and multiplexing gains.

We use Monte Carlo simulation to evaluate the average end-to-end sum-rates and multiplexing gains achievable through the proposed algorithms. First, the numerical results confirm the convergence of the proposed algorithms as expected. Second, over the iterations of the total leakage minimization algorithm, the true interference dominates at the beginning but is canceled quickly; after that, the enhanced noise from the relays becomes dominant. This means that relay-aided interference alignment should take into account the enhanced noise from the relays. Third, the total leakage minimization algorithm achieves lower end-to-end sum-rates than the others at low-to-medium SNR values because it ignores the desired signal power and noise power at the receivers. Nevertheless, the MSE-based algorithms result in unfairness, i.e., some users have much smaller rates than the others. Thus, the MSE-based algorithms achieve lower end-to-end sum-rates and multiplexing gains than the total leakage minimization algorithm at high SNR. One reason for this is that the MSE-based algorithms may either turn off some data streams or nullify the desired signals to some receivers. Fourth, for fixed numbers of antennas at the transmitters and at the receivers, even with half-duplex loss, AF relays can provide larger end-to-end multiplexing gains than DF relays or direct transmissions. Finally, the results show that AF relays provide larger average achievable end-to-end sum-rates than do DF relays. The proposed algorithms also provide higher achievable end-to-end sum-rates than the existing AF relaying strategies that do not align interference at the receivers.

The organization of the remainder of this paper is as follows. Section~\ref{sec:systemModel} describes the system model. Section~\ref{sec:problemFormulation} formulates the end-to-end sum-rate maximization problems and presents our proposed approach. Section~\ref{sec:algorithms} develops three cooperative algorithms that aim at finding high-quality solutions of the sum-rate maximization problems. Section~\ref{sec:Simulations} evaluates numerically the proposed algorithms. Section~\ref{sec:Conclusions} concludes this paper and suggests future research.

Notation: We use normal letters (e.g., $a$) for scalars, lowercase and uppercase boldface letters (e.g., $\bh$ and $\bH$) for column vectors and matrices. $\bI_{N}$ and $\b0_{N}$ are the identity matrix and all-zero matrices of size $N \times N$. $\nu^{n}_{\min}(\bA)$ gives the eigenvectors corresponding to the $n$ smallest eigenvalues of $\bA$. For a matrix $\bA$, $\bA^{T}$ is the transpose matrix, $\|\bA\|^{2}_{F}$ the Frobenious norm, $\bA^{*}$ the conjugate transpose, and $\tr(\bA)$ the trace. $\vect(\bA)$ denotes the $\vect$ operator to transform $\bA$ into $\ba$ while $\vect^{-1}(\ba)$ denotes the inverse operator. $\otimes$ is the Kronecker product. $\bbE[\cdot]$ is the statistical expectation operator. $()^{(n)}$ denotes iteration index. $()_{\rT}$ is used for transmitters' parameters, $()_{\rR}$ for receivers', and $()_{\rX}$ for relays'.

\section{System Model}\label{sec:systemModel}
\begin{figure}[h]
\centering
\vspace{-5pt}
\includegraphics[width=3.0in]{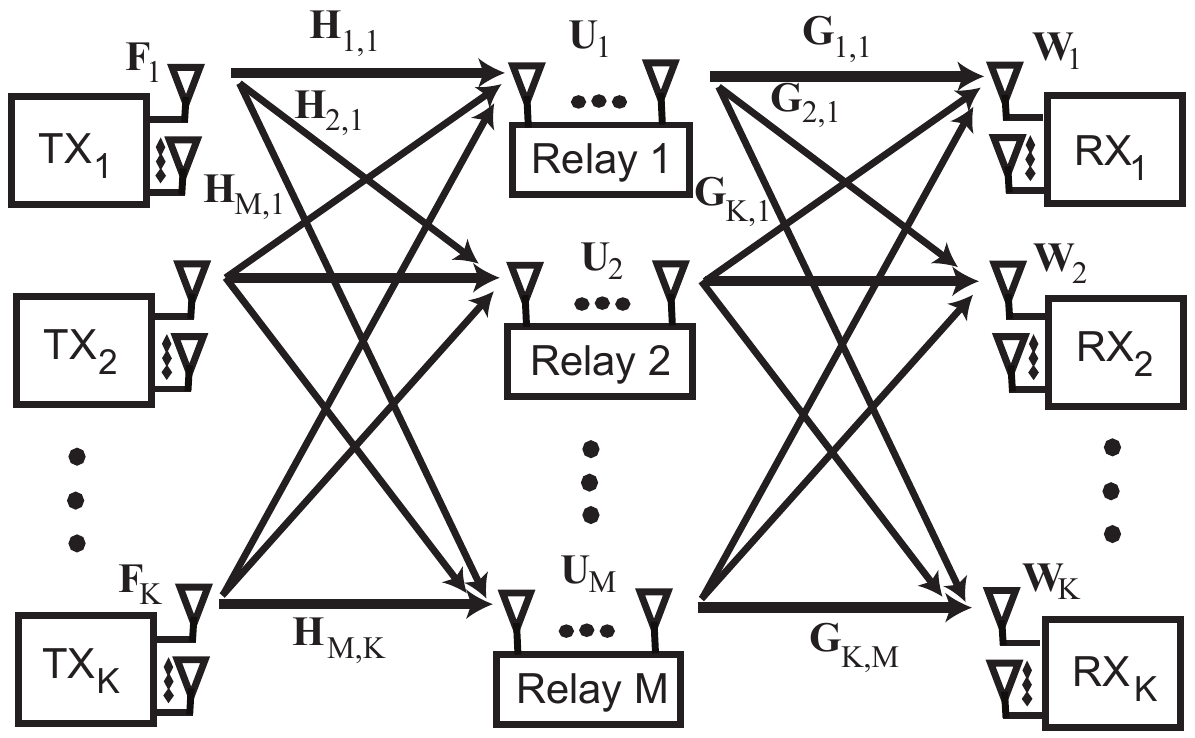}
\caption{A relay interference channel where $M$ half-duplex AF relays aid the communication of $K$ transmitter-receiver pairs.}
\label{fig:SystemDiagram}
\vspace{-15pt}
\end{figure}

Consider a relay interference channel where $M$ half-duplex AF relays aid the one-way communication between $K$ pairs of transmitters and receivers, as illustrated in Fig. \ref{fig:SystemDiagram}. Each transmitter has data for only one receiver and each receiver is served by only one transmitter. Each pair is assigned a unique index $k \in \cK\triangleq \{1, \cdots, K\}$. Transmitter $k$ has $N_{\rT,k}$ antennas while receiver $k$ has $N_{\rR,k}$ antennas for $k \in \cK$. Similarly, each relay is assigned a unique index $m \in \cM \triangleq \{1, \cdots, M\}$. Relay $m$ has $N_{\rX,m}$ antennas for $m\in \cM$. The half-duplex relays cannot transmit and receive at the same time, thus the transmission procedure consists of two stages. In the first stage, the transmitters send data to the relays. In the second stage, the relays apply linear processing to the received signals and forward to the receivers. We assume the direct channels between the transmitters and the receivers are ignored by the second-stage receivers.

We denote $\bH_{m,k} \in \bbC^{N_{\rX,m} \times N_{\rT,k}}$ as the matrix channel from transmitter $k$ to relay $m$ and $\bG_{k,m} \in \bbC^{N_{\rR,k} \times N_{\rX,m}}$ as the matrix channel from relay $m$ to receiver $k$ for $k \in \cK$ and $m \in \cM$. We assume that perfect and instantaneous knowledge of $\bH_{m,k}$ and $\bG_{k,m}$ for $k \in \cK$ and $m \in \cM$ is available at a central processing unit. Although this is a strict requirement, our results are still valuable since they show the substantial gains that can be achieved through coordination. Our results can be used as a benchmark for future work that makes more practical CSI assumptions. 

Let $\bs_{k} \in \bbC^{d_{k} \times 1}$ be the transmit symbol vector at transmitter $k$, where $d_{k} \leq \min\{N_{\rT,k},N_{\rR,k}\}$ is the number of data streams from transmitter $k$ to receiver $k$ for $k \in \cK$. The transmit symbols are independent identically distributed (i.i.d.) such that $\bbE(\bs_{k}\bs^{*}_{k})=\bI_{d_{k}}$. Transmitter $k$ uses a linear transmit precoder $\bF_{k} \in \bbC^{N_{\rT,k} \times d_{k}}$ to map $\bs_{k}$ to its transmit antennas. Let  $p^{\max}_{\rT,k}$ be the maximum transmit power. The actual transmit power at transmitter $k$ is $p_{\rT,k} = \tr(\bF^{*}_{k}\bF_{k})$. Let $\bn_{\rX,m}$ be spatially white, additive Gaussian noise at relay $m$ with covariance $\bbE(\bn_{\rX,m}\bn^{*}_{\rX,m})=\sigma^{2}_{\rX,m}\bI_{N_{\rX,m}}$ for $m\in\cM$. With perfect synchronization, relay $m$ observes the following signal
\begin{eqnarray}
\by_{\rX,m} = \sum_{k=1}^{K} \underbrace{\bH_{m,k}\bF_{k}}_{\bcH_{m,k}}\bs_{k}+\bn_{\rX,m}.
\end{eqnarray}
Let $\bU_{m} \in \bbC^{N_{\rX,m} \times N_{\rX,m}}$ be the processing matrix at relay $m$. The transmit signal at relay $m$ is given by
\begin{eqnarray}
\bx_{\rX,m} = \bU_{m}\by_{\rX,m} = \sum_{k=1}^{K} \bU_{m}\bcH_{m,k}\bs_{k}+\bU_{m}\bn_{\rX,m}.
\end{eqnarray}
Relay $m$ actually uses the following transmit power
\begin{eqnarray}
p_{\rX, m} = \sum_{k=1}^{K} \tr(\bU_{m}\bcH_{m,k}\bcH^{*}_{m,k}\bU^{*}_{m})+ \sigma_{\rX,m}^{2}\tr(\bU_{m}\bU^{*}_{m}).\label{eq:actualRelayPower}
\end{eqnarray}

There are two possible types of power constraints at the relays: i) a set of individual power constraints at the relays and ii) a sum power constraint at all the relays. Individual relay power constraints are often considered in the cellular system literature~\cite{RenSchaar2010:TWC, RamanEtAl2011:TWC}. While a sum power constraint is often considered in the ad hoc network literature to extend the lifetime of battery-powered relays~\cite{JafarEtAl2007:TIT,LiangSchober2010:VTC}. Let $p^{\max}_{\rX,m}$ be the maximum transmit power at relay $m$ and $p^{\max}_{\rX}$ be the maximum sum transmit power at all the relays. When power control is considered, the individual relay power constraints are 
$p_{\rX, m} \leq p^{\max}_{\rX, m}, \forall m \in \cM$; whereas the sum relay power constraint is $\sum_{m=1}^{M} p_{\rX, m} \leq p^{\max}_{\rX}$. Without power control, the inequalities in the power constraint expressions are replaced by equalities. The following sections focus on the sum power constraint at the relays. Section \ref{subsec:discussion} discusses the applicability of individual relay power constraints while Section V simulates the impact of individual relay power constraints on achievable end-to-end sum-rates.

Let $\bn_{\rR,k}$ be spatially white, additive Gaussian noise at receiver $k$ with covariance $\bbE(\bn_{\rR,k}\bn^{*}_{\rR,k}) = \sigma^{2}_{\rR,k}\bI_{N_{\rR,k}}$. We denote $\bcG_{k,m} = \bG_{k,m}\bU_{m}$. Receiver $k$ observes the following signal
\begin{eqnarray}
\by_{k} &=& \sum_{m=1}^{M} \bG_{k,m}\bx_{\rX,m} + \bn_{\rR,k}\label{eq:receivedSignal1}\\
&=& \sum_{q=1}^{K} \underbrace{\sum_{m=1}^{M}\bcG_{k,m}\bcH_{m,q}}_{\bcT_{k,q}}\bs_{q} + \sum_{m=1}^{M}\bcG_{k,m}\bn_{\rX,m}+\bn_{\rR,k}\label{eq:receivedSignal2},
\end{eqnarray}
where $\bcT_{k,q}$ is the effective end-to-end channel from transmitter $q$ to receiver $k$ for $k, q \in \cK$. Applying a linear receive filter $\bW_{k} \in \bbC^{N_{\rR,k} \times d_{k}}$ to $\by_{k}$, receiver $k$ obtains
\begin{eqnarray}\label{eq:postProcessingSignal}
\bar{\by}_{k} 
&=& \underbrace{\bW^{*}_{k}\bcT_{k,k}\bs_{k}}_{\mathrm{desired~signal}} + \underbrace{\sum_{\substack{q=1\\q\neq k}}^{K}\bW^{*}_{k}\bcT_{k,q}\bs_{q}}_{\mathrm{interference}} + \underbrace{\sum_{m=1}^{M}\bW^{*}_{k}\bcG_{k,m}\bn_{\rX,m}}_{\mathrm{enhanced~noise~from~relays}}+ \underbrace{\bW^{*}_{k}\bn_{\rR,k}}_{\mathrm{local~noise}}.
\end{eqnarray}
The pre-processing interference-plus-noise covariance matrix at receiver $k$ is 
\begin{eqnarray}
\bR_{k} &=& \sum_{\substack{q=1\\q\neq k}}^{K} \bcT_{k,q}\bcT^{*}_{k,q} + \sum_{m=1}^{M} \sigma^{2}_{\rX,m}\bcG_{k,m}\bcG^{*}_{k,m}+\sigma^{2}_{\rR,k}\bI_{d_{k}}.\label{eq:NoiseCovMat}
\end{eqnarray}

For notational convenience, we denote $\{\bF\} \triangleq \{\bF_{k}\}_{k=1}^{K}$, $\{\bU\} \triangleq \{\bU_{m}\}_{m=1}^{M}$ and $\{\bW\} \triangleq \{\bW_{k}\}_{k=1}^{K}$. We also denote $\bcU_{m,k}\triangleq \bU_{m}\bH_{m,k}$ and $\bcW_{k,m} \triangleq \bW^{*}_{k}\bG_{k,m}$ for $k \in \cK$ and $m \in \cM$. Table \ref{tab:equivChannel} summarizes the notation of equivalent channel gain matrices used in the paper for $k,q \in \cK$ and $m \in \cM$.
\begin{table}[!h]
\centering
\caption{Notation of equivalent channel matrices used in the paper for $k,q \in \cK$ and $m \in \cM$.}
\begin{tabular}{| l | l |}
\hline
Equivalent channel matrix & Definition \\
\hline
$\bcH_{m,k}$ & $\bH_{m,k}\bF_{k}$\\
\hline
$\bcG_{k,m}$ & $\bG_{k,m}\bU_{m}$\\ 
\hline
$\bcU_{m,k}$ & $\bU_{m}\bH_{m,k}$\\
\hline
$\bcW_{k,m}$ & $ \bW^{*}_{k}\bG_{k,m}$\\
\hline
$\bcT_{k,q}$ & $\sum_{m=1}^{M}\bcG_{k,m}\bcH_{m,q} = \sum_{m=1}^{M} \bG_{k,m}\bU_{m}\bH_{mq}\bF_{q}$\\
\hline
\end{tabular}
\label{tab:equivChannel}
\end{table}

\section{Problem Formulation and Proposed Approach}\label{sec:problemFormulation}
We formulate the end-to-end sum-rate maximization problem in Section~\ref{subsec:sumrateMax} and propose an approach to solving it in Section~\ref{subsec:proposedApproach}.

\subsection{End-to-end Sum-Rate Maximization}\label{subsec:sumrateMax}
For tractable analysis, we assume Gaussian signaling is used. For a given $\{\bF\}$ and $\{\bU\}$, the achievable rate for the $k$-th transmitter-receiver pair is maximized by using the linear MMSE receive filter~\cite{McKayEtAl2010:TIT}
\begin{eqnarray}
\bW^{\rMMSE}_{k} = (\bcT_{k,k}\bcT^{*}_{k,k}+\bR_{k})^{-1}\bcT_{k,k},\label{eq:MMSEreceiver}
\end{eqnarray}
where $\bR_{k}$ is given in (\ref{eq:NoiseCovMat}). Thus, we only need to focus on the design of $\{\bF\}$ and $\{\bU\}$. Note that the corresponding maximum achievable rate is given by~\cite{McKayEtAl2010:TIT}
\begin{eqnarray}
R_{k}\big(\{\bF\},\{\bU\}\big) &=& \log_{2}\det \big(\bI_{d_{k}} + \bcT^{*}_{k,k}\bR^{-1}_{k}\bcT_{k,k}\big).
\end{eqnarray}
The sum of the end-to-end achievable rates is defined as 
\begin{eqnarray}
R_{\rsum}\big(\{\bF\},\{\bU\}\big) 
&=&  - \sum_{k=1}^{K} \log_{2}\det\Big(\bE^{\rMMSE}_{k}(\{\bF\},\{\bU\})\Big).\label{eq:Rsum}
\end{eqnarray}
The end-to-end sum-rate maximization problem without power control is formulated as follows
\begin{eqnarray}
(\cOP\mbox{-}\rnoPC):~\min_{\{\bF\},\{\bU\}} 
&& -R_{\rsum}\big(\{\bF\},\{\bU\}\big)\nonumber\\
\mbox{s.t.} && p_{\rT,k} = p^{\rmax}_{\rT,k}, k=1, \cdots, K,\label{eq:SR1}\\
&& \sum_{m=1}^{M} p_{\rX,m} = p^{\rmax}_{\rX}.\label{eq:SR2}
\end{eqnarray}

\begin{Remark}
The counterpart problem with power control can be obtained by replacing the equalities in the constraints by the inequalities. We denote it as $(\cOP\mbox{-}\rPC)$. Power control may improve the end-to-end sum-rates at the expense of additional overhead because the central unit needs to inform the transmitters about both the norm and the shape of designed transmit precoders.
\end{Remark}

\begin{Remark}\label{remark:feasibility}
The following $(\bF_{0,k},\bW_{0,k},\bU_{0,m})$ for $k \in \cK$ and $m \in \cM$ satisfies the constraints of both ($\cOP\mbox{-}\rPC$) and $(\cOP\mbox{-}\rnoPC)$
\begin{eqnarray}
\bF_{0,k} &=& \sqrt{\frac{p^{\max}_{\rT,k}}{d_{k}}}\bI_{N_{\rT,k} \times d_{k}}, k \in \cK,\label{eq:FeasibleF}\\
\bW_{0,k} &=& \sqrt{\frac{1}{d_{k}}}\bI_{N_{\rR,k} \times d_{k}}, k \in \cK,\label{eq:FeasibleW}\\
\bU_{0,m} &=&\sqrt{\alpha p^{\max}_{\rX}}\bI_{N_{\rX,m} \times N_{\rX,m}}, m \in \cM,\label{eq:FeasibleU}
\end{eqnarray}
where $\alpha = \Big(\sum_{k=1}^{K} \frac{p^{\max}_{\rT,k}}{d_{k}}\sum_{m=1}^{M}\tr(\bI_{d_{k} \times N_{\rT,k}}\bH^{*}_{m,k}\bH_{m,k}\bI_{N_{\rT,k} \times d_{k}})+\sum_{m=1}^{M}N_{\rX,m}\sigma^{2}_{\rX,m}\Big)^{-1}$.
\end{Remark}

\begin{Remark}
($\cOP\mbox{-}\rPC$) and $(\cOP\mbox{-}\rnoPC)$ are nonconvex and NP-hard. Moreover, even the smallest configuration of the MIMO AF relay interference channel with $K = M = 2$ and $N_{\rT} = N_{\rX} = N_{\rR} = 2$ requires the determination of twelve complex variables for the transmit precoders and relay processing matrices, which makes even a brute force approach challenging. 
\end{Remark}

\subsection{Proposed Approach}\label{subsec:proposedApproach}
Instead of directly solving for the globally optimal solutions of ($\cOP\mbox{-}\rPC$) and $(\cOP\mbox{-}\rnoPC)$, we aim at finding their high-quality solutions with reasonable computational complexity. To do this, in Section \ref{subsec:TLformulation} and Section \ref{subsec:SRformulation}, we formulate two classes of new optimizations problems that have exactly the same constraints as ($\cOP\mbox{-}\rPC$) or $(\cOP\mbox{-}\rnoPC)$ but with different objective functions.

\subsubsection{Total Leakage Minimization}\label{subsec:TLformulation}
This section presents an approach for interference alignment in the AF relay interference channel, which is inspired by those for the single-hop interference channel in~\cite{PetersHeath2008:ICASSP, PetersHeath2011:TVT, GomadamEtAl2011:TIT}. The underlying observation for this approach is that when interference alignment is feasible, the sum power of the interference at all the receivers, also known as the leakage, is zero. From (\ref{eq:postProcessingSignal}), there are three groups of unwanted signals at each receiver: i) interference, ii) enhanced noise from the relays, and iii) local noise. We denote $\cI \left(\{\bF\},\{\bU\},\{\bW\}\right)$ as the total leakage power of the AF relay interference channel. By evaluating the expectation and exploiting the independence of transmit signals $\bs_{k}$ for $k \in \cK$ and using the equality $\|\bA\|^{2}_{F} = \tr(\bA\bA^{*})$, we obtain
\begin{eqnarray}
\cI \left(\{\bF\},\{\bU\},\{\bW\}\right) 
&=& \sum_{k=1}^{K}\sum_{\substack{q=1\\ q\neq k}}^{K} \tr(\bW^{*}_{k}\bcT_{k,q}\bcT^{*}_{k,q}\bW_{k}).\label{eq:TotalLeakage4}
\end{eqnarray}

In our opinion, the high SNR regime of the relay interference channel corresponds to high transmit power at both the transmitters and the relays. As a result, in addition to eliminating completely interference, we also need to eliminate the enhanced relay noise; otherwise, the enhanced relay noise power scales with the desired signal power, preventing the system from achieving high multiplexing gain. We denote $\cN(\{\bU\}, \{\bW\})$ as the sum power of enhanced noise from the relays. By evaluating the expectations and exploiting the independence of the noise vectors at the relays, we obtain
\begin{eqnarray}
\cN(\{\bU\}, \{\bW\})
&=& \sum_{k=1}^{K}\sum_{m=1}^{M} \sigma^{2}_{\rX,m}\tr(\bW^{*}_{k}\bcG_{k,m}\bcG^{*}_{k,m}\bW_{k}).
\end{eqnarray}

Note that scaling down transmit power at either the transmitters or relays decreases the total leakage power at the receivers. For example, if $(\{(1/a)\bF\},\{\bU\},\{\bW\})$ is used instead of $(\{\bF\},\{\bU\},\{\bW\})$ where $\{(1/a)\bF\} = \{(1/a)\bF_{1}, \cdots, (1/a)\bF_{K}\}$ and  $a >1$, then both the actual transmit power at the transmitters and the total leakage power decrease $a^{2} > 1$ times. Thus, equality power constraints at the transmitters and relays are required to obtain a meaningful design problem. This means that power control should not be considered in the context of total leakage power minimization. In other words, we do not use the total leakage minimization approach to find solutions to $(\cOP\mbox{-}\rPC)$. Also to obtain a meaningful design problem, we add the orthonormal constraints on $\bW_{k}$ as $\bW^{*}_{k}\bW_{k} = \bI_{d_{k}}$ for $k \in \cK$. Without such constraints, we can always use zero matrices as the receive filters to get zero total leakage power. Consequently, to find high-quality solutions of $(\cOP\mbox{-}\rnoPC)$, we propose to solve the following problem
\begin{eqnarray}
(\cTL):~\min_{\{\bF\},\{\bU\},\{\bW\}} && \cI\left(\{\bF\},\{\bU\},\{\bW\}\right) + \cN (\{\bU\}, \{\bW\})\nonumber\\
\mbox{s.t.}
&& p_{\rT,k} = p^{\max}_{\rT,k}, k \in \cK,\label{eq:IAPoweConstraint1}\\
&&\sum_{m=1}^{M}p_{\rX,m} = p^{\max}_{\rX},\label{eq:IAPoweConstraint3}\\
&& \bW_{k}^{*}\bW_{k} = \bI_{d_{k}}, k \in \cK\label{eq:IAPoweConstraint4}.
\end{eqnarray}
Note that ($\cTL$) is nonconvex and in general is NP-hard. Also, $(\cTL)$ does not take into account the desired signal power  and local noise at the receivers.

\begin{Remark}\label{remark:relatedRelayIApriorWork}
The total leakage minimization problem formulated in~\cite{NingEtAl2010:ISIT} for an AF relay network is a simplified version of ($\cTL$). It is assumed in~\cite{NingEtAl2010:ISIT} that the transmitters and receivers are equipped with a single antenna. Each pair is aided by a dedicated multiple-antenna AF relay. The formulation in~\cite{NingEtAl2010:ISIT} does not consider power constraints at the relays. In addition, it is assumed that there are no cross-links for the transmissions from relays to receivers, i.e. $\bG_{k,q} =\b0$ for all $k, q \in \cK$ and $k \neq q$. As a result, for fixed $\{\bF\}$ and $\{\bW\}$, the algorithm in~\cite{NingEtAl2010:ISIT} can determine each $\bU_{m}$ separately. 
\end{Remark}

\subsubsection{Sum Mean Squared Error Minimization}\label{subsec:SRformulation}
The section presents another approach that is based on a relationship between the achievable rates and MSE values at the receivers with Gaussian signaling~\cite{GuoEtAl2005:TIT}. This is inspired by prior work on the MIMO broadcast channel~\cite{ChristensenEtAl2008:TWC}, MIMO interference channel~\cite{SchmidtEtAl2009:Asilomar}, MIMO interference broadcast channel~\cite{RazaviyaynEtAl2011:CISS}, and two-way relay channel~\cite{XuHua2010:ICASSP,LeeEtAl2010:TWC}. Let $\bE_{k}(\{\bE\},\{\bU\},\bW_{k})$ be the MSE matrix at receiver $k$. After some manipulation, we obtain
\begin{eqnarray}
\bE_{k}(\{\bE\},\{\bU\},\bW_{k}) = \bW^{*}_{k}(\bcT_{k,k}\bcT^{*}_{k,k}+\bR_{k})\bW_{k} - \bW^{*}_{k}\bcT_{k,k} - \bcT^{*}_{k,k}\bW_{k} + \bI_{d_{k}}.\label{eq:Ek}
\end{eqnarray}
Note that the MSE at receiver $k$, defined as $MSE_{k} = \tr(\bE_{k}(\{\bE\},\{\bU\},\bW_{k}))$, is minimized by the linear MMSE receive filter $\bW^{\rMMSE}_{k}$. Moreover, it is well-established that~\cite{GuoEtAl2005:TIT,ChristensenEtAl2008:TWC,SchmidtEtAl2009:Asilomar,XuHua2010:ICASSP,LeeEtAl2010:TWC,RazaviyaynEtAl2011:CISS,ShiEtAl2011:TSP}
\begin{eqnarray}
R_{k}\big(\{\bF\},\{\bU\}\big) = -\log_{2}\det(\bE_{k}(\{\bF\},\{\bU\},\bW^{\rMMSE}_{k})).
\end{eqnarray}

We introduce auxiliary weight matrix variables $\{\bV\} \triangleq (\bV_{1},\cdots,\bV_{K})$ that are square ($\bV_{k} \in \bbC^{d_{k} \times d_{k}}$) and positive semidefinite for $k \in \cK$. The weight matrices $\bV_{k}$ are just auxiliary variables for the optimization technique and have no actual physical meaning. Define the matrix-weighted sum of MSE values as follows
\begin{eqnarray}
WMSE_{\rsum}(\{\bF\},\{\bU\},\{\bW\},\{\bV\}) = \sum_{k=1}^{K} \bigg(\tr\Big(\bV_{k}\bE_{k}\big(\{\bF\},\{\bU\},\{\bW\}\big)\Big) - \log_{2}\det\big(\bV_{k}\big)\bigg).\label{eq:WMSEsum}
\end{eqnarray}
Then, we formulate the following weighted sum-MSE minimization problem
\begin{eqnarray}
(\cWMSE\mbox{-}\rnoPC):~\min_{\{\bF\},\{\bU\},\{\bW\},\{\bV\}} && WMSE_{\rsum}\big(\{\bF\},\{\bU\},\{\bW\},\{\bV\}\big)\label{eq:WMMSE1}\nonumber\\
\mbox{s.t.} 
&& p_{\rT,k} = p^{\rmax}_{\rT,k}, k \in \cK \label{eq:WMMSE3}\\
&& \sum_{m=1}^{M} p_{\rX,m} = p^{\rmax}_{\rX}.\label{eq:WMMSE4}
\end{eqnarray}
Similarly, we formulate $(\cWMSE\mbox{-}\rPC)$ by replacing the equalities in $(\cWMSE\mbox{-}\rnoPC)$. Using the same steps in~\cite{RazaviyaynEtAl2011:CISS,ShiEtAl2011:TSP}, we can show that $(\cWMSE\mbox{-}\rnoPC)$ and $(\cOP\mbox{-}\rnoPC)$ have exactly the same stationary points if we use the linear MMSE receivers and choose the following matrix weights
\begin{eqnarray}
\bV^{\ropt}_{k}\big(\{\bF\},\{\bU\},\bW^{\rMMSE}_{k}\big) &=& \bE^{-1}_{k}\big(\{\bF\},\{\bU\},\bW^{\rMMSE}_{k}),\label{eq:Vopt}\\
& = & \bI_{d_{k}} + \bcT^{*}_{k,k} \bR^{-1}_{k}\bcT_{k,k}\label{eq:VoptMMSE}.
\end{eqnarray}
This observation is also true for $(\cWMSE\mbox{-}\rPC)$ and $(\cOP\mbox{-}\rPC)$. Thus, instead of directly solving $(\cOP\mbox{-}\rnoPC)$ (or $(\cOP\mbox{-}\rPC)$), we can focus on finding high-quality solutions to its corresponding weighted sum-MSE minimization problem, which has a better-behaved objective function.

\begin{Remark}\label{remark:WMSEconvexity}
The matrix-weighted sum-MSE value $WMSE_{\rsum}(\{\bF\},\{\bU\},\{\bW\},\{\bV\})$ is convex with respect to $\bF_{k}$ for $k \in \cK$ if we always choose $\bV_{k}$ according to (\ref{eq:Vopt}). Indeed, we can check that $MSE_{k}$ is convex with respect to $\bF_{q}$ for all $k, q \in \cK$. By construction, $\bV^{\ropt}_{k}\big(\{\bF\},\{\bU\},\bW^{\rMMSE}_{k}\big)$ is a Hermitian and positive semidefinite matrix for $k \in \cK$. Then, by definition $WMSE_{\rsum}(\{\bF\},\{\bU\},\{\bW\},\{\bV\})$ is also convex with respect to $\bF_{k}$ for $k \in \cK$.
\end{Remark}

\section{Algorithms}\label{sec:algorithms}
The problems $(\cTL)$, $(\cWMSE\mbox{-}\rnoPC)$, and $(\cWMSE\mbox{-}\rPC)$ formulated in Section \ref{subsec:proposedApproach} are nonconvex and in general are NP-hard. In this section, rather than attempting solving for their globally optimal solutions, we adopt an alternating minimization approach~\cite{Bertsekas1999:BOOK} to develop iterative algorithms for finding their high-quality solutions. In each iteration, we alternatively fix all but one variable and determine the remaining variable by solving a single-variable optimization problem. The optimization problem in each iteration is always feasible since it has the outcome of the previous iteration as a feasible point. After initialization, the algorithms are repeated until a convergent point is reached. Section \ref{subsec:TLalgorithm} presents the algorithm for solving ($\cTL$), which is denoted as Algorithm 1. Two algorithms for solving $(\cWMSE\mbox{-}\rnoPC)$ and $(\cWMSE\mbox{-}\rPC)$ are presented in Section \ref{subsec:SRnoPCalgorithm} and Section \ref{subsec:SRPCalgorithm}. They are denoted as Algorithm 2 and Algorithm 3, respectively.

\subsection{Algorithm for Total Leakage Minimization $(\cTL)$}\label{subsec:TLalgorithm}
There are three classes of design subproblems in Algorithm 1: i) receiver filter design, ii) relay processing matrix design, and iii) transmit precoder design. 

\subsubsection{Receive Filter Design for ($\cTL$)}\label{subsection:TLreceiveFilterDesign}
We can rewrite the cost function as $\sum_{k=1}^{K} \tr(\bW^{*}_{k}\bZ_{k}\bW_{k})$, where $\bZ_{k} = \sum_{\substack{q=1\\q\neq k}}^{K}\bcT_{k,q}\bcT^{*}_{k,q} + \sum_{m=1}^{M}\sigma^{2}_{\rX,m}\bcG_{k,m}\bcG^{*}_{k,m}$. Since $\bW_{k}$ for $k \in \cK$ are decoupled in the cost function, they can be determined separately and in parallel by solving
\begin{eqnarray}
(\cTL\mbox{-}\bW_{k}): \bW_{k} = \argmin_{\bX \in \bbC^{N_{\rR,k} \times d_{k}}: \bX^{*}\bX=\bI_{d_{k}}} \tr(\bX^{*}\bZ_{k}\bX)\label{eq:WOP}.\nonumber
\end{eqnarray}
It follows from~\cite{Lutkepohl1996:BOOK} that a global optimum of $(\cTL\mbox{-}\bW_{k})$ is $\bW_{k} = \nu^{d_{k}}_{\min}(\bZ_{k})$.

\subsubsection{Relay Processing Matrix Design for $(\cTL)$}\label{subsec:TLrelayMatrixDesign}
We focus on determining $\bU_{m}$ for some $m \in \cM$ by solving the following single-variable optimization problem $(\cTL\mbox{-}\bU_{m})$
\begin{eqnarray}
\min_{\bX \in \bbC ^{N_{\rX,m} \times N_{\rX, m}}} && \sum_{k=1}^{K}\sum_{\substack{q=1\\\neq k}}^{K} \tr\Big(\bX\bcH_{m,q}\bcH^{*}_{m,q}\bX^{*}\bcW^{*}_{k,m}\bcW_{k,m}\Big) + \sigma^{2}_{\rX,m}\sum_{k=1}^{K} \tr(\bX^{*}\bcW^{*}_{k,m}\bcW_{k,m}\bX)\nonumber\\
&& +~\sum_{k=1}^{K}\sum_{\substack{q=1\\q\neq k}}^{K} \sum_{\substack{n=1\\n\neq m}}^{M} \tr\bigg(\bX\bcH_{m,q}\bcH^{*}_{n,q}\bU^{*}_{n}\bcW^{*}_{k,n}\bcW_{k,m}\bigg)\nonumber\\
&& +~\sum_{k=1}^{K}\sum_{\substack{q=1\\q\neq k}}^{K} \sum_{\substack{n=1\\n\neq m}}^{M} \tr\bigg(\bcW_{k,m}^{*}\bcW_{k,n}\bU_{n}\bcH_{n,q}\bcH^{*}_{m,q}\bX^{*}\bigg)\nonumber\\
\mbox{s.t.} && \tr\bigg(\bX\Big(\sum_{k=1}^{K}\bcH_{m, k}\bcH^{*}_{m,k} + \sigma^{2}_{\rX,m}\bI_{N_{\rX,m}}\Big)\bX^{*}\bigg) = \eta_{\rU,m},\label{eq:TLUmConstraint}
\end{eqnarray}
where  
$\eta_{\rU,m} =p^{\max}_{\rX} - \sum_{\substack{n=1\\ n \neq m}}^{M}\sum_{k=1}^{K} \tr\left(\bU_{n}\bcH_{n, k}\bcH^{*}_{n,k}\bU^{*}_{n}\right) -  \sum_{\substack{n=1\\ n \neq m}}^{M}\sigma^{2}_{\rX,n} \tr(\bU_{n}\bU^{*}_{n})$.
Because of the special form of the first term in the cost function of $(\cTL\mbox{-}\bU_{m})$, it is not straightforward to use the methods for the single-hop interference channel like those in \cite{PetersHeath2011:TVT} to solve $(\cTL\mbox{-}\bU_{m})$.

We propose to transform $(\cTL\mbox{-}\bU_{m})$ into a more readily solvable form by introducing a new variable $\bu_{m} = \vect(\bU_{m}) \in \bbC^{N^{2}_{\rX,m} \times 1}$. We define the following matrices that are independent of $\bu_{m}$
\begin{eqnarray}
\bA_{1,m}&=&\sum_{k=1}^{K} \bigg(\sum_{\substack{q=1\\q\neq k}}^{K} \bcH_{m,q}\bcH^{*}_{m,q}+ \sigma^{2}_{\rX,m}\bI_{N_{\rX,m}}\bigg)^{T}\otimes\left(\bcW^{*}_{k,m}\bcW_{k,m}\right),\label{eq:TLU1}\\
\ba_{2,m} &=& \vect\bigg(\sum_{k=1}^{K}\sum_{\substack{q=1\\q\neq k}}^{K} \sum_{\substack{n=1\\n\neq m}}^{M} \bcW^{*}_{k,m}\bcW_{k,n}\bU_{n} \bcH_{n,q}\bcH^{*}_{m,q}\bigg),\label{eq:TLU2}\\
\bA_{3,m} &=& \bigg(\sum_{k=1}^{K}\bcH_{m, k}\bcH^{*}_{m,k} + \sigma^{2}_{\rX,m}\bI_{N_{\rX,m}}\bigg)^{T} \otimes \bI_{N_{\rX,m}}.\label{eq:TLU3}
\end{eqnarray}
Note that with probability one, $\bA_{3,m}$ is Hermitian and positive definite while $\bA_{1,m}$ is Hermitian and positive semidefinite. Then, we use the following equalities, $\tr(\bA\bB\bA^{*}\bC)=(\vect(\bA))^{*}(\bB^{T} \otimes \bC)\vect(\bA)$, $\tr(\bA^{*}\bB\bA) = \tr(\bA\bI\bA^{*}\bB) = (\vect(\bA))^{*}(\bI \otimes \bB)\vect(\bA)$ and $\tr(\bA\bB^{*}) = (\vect(\bB))^{*}\vect(\bA)$~\cite{HornJohnson1991:BOOK}, to transform both the cost function and the constraint of $(\cTL\mbox{-}\bU_{m})$ into quadratic expressions of $\bu_{m}$. The quadratically constrained quadratic program (QCQP) for designing $\bu_{m}$ is
\begin{eqnarray}
(\cTL\mbox{-}\bu_{m}):~\min_{\bx \in \bbC^{N^{2}_{\rX,m} \times 1}} && \bx^{*}\bA_{1,m}\bx + \ba^{*}_{2,m}\bx + \bx^{*}\ba_{2,m}\nonumber\\
\mbox{s.t.} && \bx^{*} \bA_{3,m}\bx = \eta_{\rU,m}\label{eq:TLU5}.
\end{eqnarray}
This is a QCQP with a single equality quadratic constraint. It is nonconvex as well.

In solving $(\cTL\mbox{-}\bu_{m})$, we introduce a new variable $\bY = \begin{pmatrix}\bu_{m} \\ 1\end{pmatrix}\begin{pmatrix}\bu_{m}^{*} & 1\end{pmatrix} = \begin{pmatrix}\bu_{m} \bu_{m}^{*} & \bu_{m} \\ \bu_{m}^{*} & 1\end{pmatrix} \in \bbC^{(N_{\rX,m}^{2} + 1) \times (N_{\rX,m}^{2} + 1)}$. It follows that $\bY$ is a rank-one Hermitian positive semidefinite matrix with bottom right entry equal to 1. We can rewrite $(\cTL\mbox{-}\bu_{m})$ equivalently as
\begin{eqnarray}
(\cTL\mbox{-}\bu_{m}\bu_{m}^{*}):~\min_{\bY \in \bbC^{(N_{\rX,m}^{2} + 1) \times (N_{\rX,m}^{2} + 1)}}&~&\tr\left(\begin{pmatrix} \bA_{1,m} & \ba_{2,m}\\ \ba_{2,m}^{*} & 1\end{pmatrix} \bY\right)\nonumber\\
\mbox{s.t.}&~& \tr\left(\begin{pmatrix} \bA_{3,m} & \b0_{N^{2}_{\rX,m} \times 1} \\ \b0_{1 \times N^{2}_{\rX,m}} & 1\end{pmatrix} \bY\right) = \eta_{\rU,m} + 1,\\
&~& \tr\left(\begin{pmatrix} \b0_{N^{2}_{\rX,m} \times N^{2}_{\rX,m}}& \b0_{N^{2}_{\rX,m} \times 1} \\ \b0_{1 \times N^{2}_{\rX,m}} & 1\end{pmatrix} \bY\right) = 1,\\
&~& \bY \succeq \b0,~\rank(\bY) = 1.
\end{eqnarray}
While the cost function and all other constraints are convex, the rank constraint is nonconvex. This rank constraint is actually the main difficulty in solving $(\cTL\mbox{-}\bu_{m}\bu_{m}^{*})$. Dropping this rank constraint, however, we obtain a relaxed version of $(\cTL\mbox{-}\bu_{m}\bu_{m}^{*})$, which is a convex optimization problem and also known as a semidefinite relaxation (SDR) of $(\cTL\mbox{-}\bu_{m}\bu_{m}^{*})$. Note that a complex-valued separable homogeneous QCQP with $n$ constraints is guaranteed to have a global optimum with rank $r \leq \sqrt{n}$~\cite{HuangPalomar2010a:TSP}. Therefore, having $n=1$ constraints, $(\cTL\mbox{-}\bu_{m}\bu_{m}^{*})$ is guaranteed to have a rank-one global optimum. The SDR of $(\cTL\mbox{-}\bu_{m}\bu_{m}^{*})$ can be solved, to any arbitrary accuracy, in a numerically reliable and efficient manner by readily available software packages, e.g., the convex optimization toolbox CVX~\cite{CVX:Software}. It is not guaranteed, however, that solving the SDR by the available software packages provides a desired rank-one global optimum of the SDR. Fortunately, we can construct a rank-one global optimum of the SDR from the resulting general-rank global optimum using the rank-reduction procedure in~\cite{HuangPalomar2010a:TSP}, which is an extension of the purification technique in~\cite{AliEtAl2009:MP}. The key idea in each step of the procedure is to modify the eigenvalues of the general-rank global optima to remove the largest eigenvalue. Each step of the procedure gives us another global optima with the same eigenvectors but with one fewer nonzero eigenvalues. We notice that the last entry of the column vector obtained by the decomposition of the rank-one global optimum~\cite{HuangZhang2007:MOR} may be a complex number with modulus of 1. By multiplying the resulting column vector with the conjugate of its last entry, we obtain a desired column vector in the form of $(\bu_{m}~1)^{T}$, which corresponds to another rank-one global optimum of  $(\cTL\mbox{-}\bu_{m}\bu_{m}^{*})$. We then use the $\vect^{-1}$ operator to get a globally optimal solution $\bU_{m}$ of $(\cTL\mbox{-}\bU_{m})$ from the resulting $\bu_{m}$.

\subsubsection{Transmit Precoder Design for ($\cTL$)}\label{subsection:TLtransmitPrecoderDesign}
We now focus on designing $\bF_{k}$ for some $k \in \cK$ by solving the following single-variable optimization problem
\begin{eqnarray}
(\cTL\mbox{-}\bF_{k}):~\min_{\bX \in \bbC^{N_{\rT, k} \times d_{k}}} && \tr\Bigg(\bX^{*}\bigg(\sum_{\substack{q=1\\q\neq k}}^{K} \sum_{m=1}^{M}\sum_{n=1}^{M}\bcU^{*}_{m,k}\bcW^{*}_{q,m}\bcW_{q,n}\bcU_{n,k}\bigg)\bX\Bigg)\nonumber\\
\mbox{s.t.} && \tr(\bX^{*}\bX) = p_{\rT, k}\label{eq:Fk1}\\
&& \tr\Bigg(\bX^{*}\Big(\sum_{m=1}^{M} \bcU^{*}_{m,k}\bcU_{m,k}\Big)\bX\Bigg) = \eta_{\rF,k},\label{eq:Fk2}
\end{eqnarray}
where
$\eta_{\rF,k} = p^{\max}_{\rX} - \sum_{\substack{q=1\\ q\neq k}}^{K} \sum_{m=1}^{M} \tr\Big(\bF^{*}_{q}\bcU^{*}_{m,q}\bcU_{m,q}\bF_{q}\Big) - \sum_{m=1}^{M} \sigma^{2}_{\rX,m}\tr\Big(\bU_{m}\bU^{*}_{m}\Big).$
Note that $(\cTL\mbox{-}\bF_{k})$ is non-convex and in general is NP-hard. Since $(\cTL\mbox{-}\bF_{k})$ has two equality constraints, the use of the Lagrange multiplier method requires a more complicated 2-D search. 

Similar to Section \ref{subsec:TLrelayMatrixDesign}, we propose a method for transforming $(\cTL\mbox{-}\bF_{k})$ into an equivalent optimization problem and for solving for its global optimum. We start by defining a new variable $\bff_{k}=\vect(\bF_{k}) \in \bbC^{N_{\rT,k}d_{k} \times 1}$. We also define the following matrices which are independent of $\bff_{k}$
\begin{eqnarray}
\bB_{1,k} &=& \bI_{d_{k}}  \otimes \bigg(\sum_{\substack{q=1\\q\neq k}}^{K} \sum_{m=1}^{M}\sum_{n=1}^{M}\bcU^{*}_{m,k}\bcW^{*}_{q,m}\bcW_{q,n}\bcU_{n,k}\bigg),\label{eq:TLF1a}\\
\bB_{2,k} &=& \bI_{d_{k}} \otimes  \bigg(\sum_{m=1}^{M} \bcU^{*}_{m,k}\bcU_{m,k}\bigg).\label{eq:TLF2a}
\end{eqnarray}
Both $\bB_{1,k}$ and $\bB_{2,k}$ are Hermitian positive definite matrices. Using $\tr(\bA^{*}\bB\bA) = (\vect(\bA))^{*}(\bI\otimes \bB)\vect(\bA)$~\cite{HornJohnson1991:BOOK}, we transform $(\cTL\mbox{-}\bF_{k})$ into the following single-variable optimization problem
\begin{eqnarray}
(\cTL\mbox{-}\bff_{k}): \min_{\bx \in \bbC^{N_{\rT,k}d_{k} \times 1}} && \bx^{*}\bB_{1,k}\bx\nonumber\\
\mbox{s.t.} && \bx^{*}\bx = p^{\max}_{\rT,k},\label{eq:TLF1}\\
&& \bx^{*}\bB_{2,k}\bx = \eta_{\rF,k}.\label{eq:TLF2}
\end{eqnarray}
Note that $(\cTL\mbox{-}\bff_{k})$ is a complex-valued homogeneous QCQP with two equality quadratic constraints. Nevertheless, $(\cTL\mbox{-}\bff_{k})$ is  still nonconvex and NP-hard~\cite{LuoEtAl2010:SPM,HuangPalomar2010a:TSP}. 

In solving $(\cTL\mbox{-}\bff_{k})$, we introduce a new variable $\bY=\bx\bx^{*}$. Note that $\bY = \bx\bx^{*}$ requires that $\bY$ be a rank-one Hermitian positive semidefinite matrix. In addition, since $\ba^{*}\bB\ba = \tr(\bB\ba\ba^{*})$ for any matrix $\bB$ and any vector $\ba$~\cite{HornJohnson1990:BOOK}, we obtain an equivalent optimization problem of $(\cTL\mbox{-}\bff_{k})$ as follows
\begin{eqnarray}
(\cTL\mbox{-}\bff_{k}\bff^{*}_{k}): \min_{\bY \in \bbC^{N_{\rT,k}d_{k} \times N_{\rT,k}d_{k}}} && \tr(\bB_{1,k}\bY)\label{eq:TLFX}\nonumber\\
\mbox{s.t.} && \tr(\bY) = p^{\max}_{\rT,k},\label{eq:TLFX1}\\
&& \tr(\bB_{2,k}\bY) = \eta_{\rF,k},\label{eq:TLFX2}\\
&& \bY \succeq \b0,~\rank(\bY) = 1.\label{eq:TLFX3}
\end{eqnarray}
Similar to solving $(\cTL\mbox{-}\bu_{m}\bu_{m}^{*})$, we adopt the SDP method for solving $(\cTL\mbox{-}\bff_{k}\bff^{*}_{k})$. Since $(\cTL\mbox{-}\bff_{k})$ is a complex-valued separable homogeneous QCQP with $n = 2$ constraints, it is guaranteed that $(\cTL\mbox{-}\bff_{k}\bff^{*}_{k})$ has a global optimum of rank $r = 1 \leq \sqrt{n}$. We can use readily available software packages, e.g., the convex optimization toolbox CVX~\cite{CVX:Software}, to solve for a general rank global optimum of the SDR of $(\cTL\mbox{-}\bff_{k}\bff^{*}_{k})$. Next, we can always construct a rank-one global optimum of the SDR from any  of its general-rank global optimum, e.g., by using the rank reduction procedure in~\cite{HuangPalomar2010a:TSP}. The decomposition of the rank-one global optimum~\cite{HuangZhang2007:MOR} gives us the desired $\bff_{k}$.  Finally, we use the $\vect^{-1}$ operator to get a globally optimal solution $\bF_{k}$ of $(\cTL\mbox{-}\bF_{k})$ from the resulting $\bff_{k}$.

\subsection{Algorithm for Sum MSE Minimization without Power Control $(\cWMSE\mbox{-}\rnoPC)$} \label{subsec:SRnoPCalgorithm}
The design subproblems in the iterations of Algorithm 2 belong to one of the following four categories.

\subsubsection{Matrix Weight Design for $(\cWMSE\mbox{-}\rnoPC)$} Since the matrix weights $\bV^{\ropt}_{k}$ for $k \in \cK$ are independent of each other, they can be updated in parallel based on (\ref{eq:Vopt}).

\subsubsection{Receive Filter Design for $(\cWMSE\mbox{-}\rnoPC)$} Recall that this approach requires the receivers use the linear MMSE receive filters $\bW^{\rMMSE}_{k}$ given in (\ref{eq:MMSEreceiver}). The receive filters can be updated in parallel.

\subsubsection{Relay Processing Matrix Design for $(\cWMSE\mbox{-}\rnoPC)$} We focus on the design of $\bU_{m}$ for some $m \in \cM$. By substituting (\ref{eq:NoiseCovMat}) and (\ref{eq:Ek}) into (\ref{eq:WMSEsum}) and removing the terms independent of $\bU_{m}$, we obtain the objective function of the design problem for $\bU_{m}$. Let $(\cWMSE\mbox{-}\rnoPC\mbox{-}\bU_{m})$ denote the design problem of $\bU_{m}$. After some manipulation and using the fact that $\bV_{k}$ is Hermitian, we obtain the formulation of $(\cWMSE\mbox{-}\rnoPC\mbox{-}\bU_{m})$ as
\begin{eqnarray}
\min_{\bX \in \bbC^{N_{\rx, m} \times N_{\rX, m}}} && \sum_{k=1}^{K}\sum_{\substack{q=1}}^{K} \tr\Big(\bX\bcH_{m,q}\bcH^{*}_{m,q}\bX^{*}\bcW^{*}_{k,m}\bV_{k}\bcW_{k,m}\Big)+\sigma^{2}_{\rX,m}\sum_{k=1}^{K}\tr(\bX^{*}\bcW^{*}_{k,m}\bV_{k}\bcW_{k,m}\bX)\nonumber\\
&& - \sum_{k=1}^{K}\tr(\bcH_{m,k}\bV^{*}_{k}\bcW_{k,m}\bX) + \sum_{k=1}^{K}\sum_{\substack{q=1}}^{K}\sum_{\substack{n=1\\ n \neq m}}^{M} \tr(\bcH_{m,q}\bcH^{*}_{n,q}\bU^{*}_{n}\bcW^{*}_{k,n}\bV_{k}\bcW_{k,m}\bX)\nonumber\\
&&- \sum_{k=1}^{K}\tr(\bX^{*}\bcW_{k,m}^{*}\bV_{k}\bcH_{m,k}^{*})  +  \sum_{k=1}^{K}\sum_{\substack{q=1}}^{K}\sum_{\substack{n=1\\ n \neq m}}^{M} \tr(\bX^{*}\bcW_{k,m}^{*}\bV_{k}^{*}\bcW_{k,n}\bU_{n}\bcH_{n,q}\bcH_{m,q}^{*})\nonumber\\
\mbox{s.t.} && \tr\bigg(\bX^{*}\Big(\sum_{k=1}^{K}\bcH_{m, k}\bcH^{*}_{m,k} + \sigma^{2}_{\rX,m}\bI_{N_{\rX,m}}\Big)\bX\bigg) = \eta_{\rU,m}.\label{eq:MSEUmConstraint}
\end{eqnarray}

Note that $(\cWMSE\mbox{-}\rnoPC\mbox{-}\bU_{m})$ differs from $(\cTL\mbox{-}\bU_{m})$ mainly due to the appearance of $\bV_{k}$ in the cost function. We introduce a new variable $\bu_{m} = \vect(\bU_{m})$ and define the following matrices
\begin{eqnarray}
\bC_{1,m} &=& \sum_{k=1}^{K}\bigg(\sum_{q=1}^{K} \bcH_{m,q}\bcH^{*}_{m,q} + \sigma^{2}_{\rX,m}\bI_{N_{\rX,m}}\bigg)^{T} \otimes \Big(\bcW^{*}_{k,m}\bV_{k}\bcW_{k,m}\Big),\label{eq:MSEU1a}\\
\bc_{2,m} &=& \vect\bigg(-\sum_{k=1}^{K}\bcW^{*}_{k,m}\bV_{k}\bcH^{*}_{m,k} + \sum_{\substack{n=1\\n\neq m}}^{M}\sum_{k=1}^{K}\sum_{q=1}^{K}\bcW^{*}_{k,m}\bV_{k}\bcW_{k,n}\bU_{n}\bcH_{n,q}\bcH^{*}_{m,q}\bigg)\label{eq:MSEU2}.
\end{eqnarray}
Using the same manipulation as in Section \ref{subsec:TLrelayMatrixDesign} and denoting $\bC_{3,m} = \bA_{3,m}$, we obtain the following equivalent optimization problem
\begin{eqnarray}
(\cWMSE\mbox{-}\rnoPC\mbox{-}\bu_{m}):~\min_{\bx \in \bbC^{N^{2}_{\rX,m} \times 1}} &&\bx^{*}\bC_{1,m}\bx + \bc^{*}_{2,m}\bx + \bx^{*}\bc_{2,m} \nonumber\\
\mbox{s.t.} &&\bx^{*}\bC_{3,m}\bx = \eta_{\rU,m}.\label{eq:MSEU1}
\end{eqnarray}
Note that $(\cWMSE\mbox{-}\rnoPC\mbox{-}\bu_{m})$ has exactly the same form as $(\cTL\mbox{-}\bu_{m})$, thus we can apply the same method used for solving $(\cTL\mbox{-}\bu_{m})$ to find a globally optimal solution $\bu_{m}$ of $(\cWMSE\mbox{-}\rnoPC\mbox{-}\bu_{m})$. We then use the $\vect^{-1}$ operator to get a globally optimal solution $\bU_{m}$ of $(\cWMSE\mbox{-}\rnoPC\mbox{-}\bU_{m})$ from the resulting $\bu_{m}$.

\subsubsection{Transmit Precoder Design for $(\cWMSE\mbox{-}\rnoPC)$} We define the following matrices for the design of $\bF_{k}$ for some $k \in \cK$
\begin{eqnarray}
\bD_{1,k} &=& \sum_{q=1}^{K}\sum_{m=1}^{M}\sum_{n=1}^{M} \bcU^{*}_{m,k}\bcW^{*}_{q,m}\bV_{q}\bcW_{q,n}\bcU_{n,k},\label{eq:MSEF1}\\
\bD_{2,k} &=& \sum_{m=1}^{M} \bcU^{*}_{m,k}\bcW^{*}_{k,m}\bV^{*}_{k},\label{eq:MSEF2}\\
\bD_{3,k} &=& \sum_{m=1}^{M} \bcU^{*}_{m,k}\bcU_{m,k}.\label{eq:MSEF4}
\end{eqnarray}
After some manipulation, we obtain the following single-variable optimization problem
\begin{eqnarray}
(\cWMSE\mbox{-}\rnoPC\mbox{-}\bF_{k}): \min_{\bX \in \bbC^{N_{\rT,k}\times d_{k}}} && \tr(\bX^{*}\bD_{1,k}\bX) - \tr(\bD^{*}_{2,k}\bX) - \tr(\bD_{2,k}\bX^{*}) \nonumber\\
\mbox{s.t.} && \tr(\bX^{*}\bX) = p^{\max}_{\rT},\label{eq:WMSEF1}\\
&& \tr(\bX^{*}\bD_{3,k}\bX) = \eta_{\rT,k}.\label{eq:WMSEF2}
\end{eqnarray}

Recall that we define $\bff_{k} = \vect(\bF_{k})$. We introduce a new variable $\bY = \begin{pmatrix}\bff_{k} \\ 1\end{pmatrix}\begin{pmatrix}\bff_{k}^{*} & 1\end{pmatrix} = \begin{pmatrix}\bff_{k}\bff_{k}^{*} & \bff_{k} \\ \bff_{k}^{*} & 1\end{pmatrix}$.
It follows that $\bY$ is a rank-one Hermitian positive semidefinite matrix with the bottom right entry equal to 1. Then, we transform $(\cWMSE\mbox{-}\rnoPC\mbox{-}\bF_{k})$ equivalently into the following problem 
\begin{eqnarray}
(\cWMSE\mbox{-}\rnoPC\mbox{-}\bff_{k}\bff^{*}_{k}):~\min_{\bY \in \bbC^{(N_{\rT,k}d_{k} +1) \times (N_{\rT,k}d_{k} +1)}} && \tr\left(\begin{pmatrix}\bI_{d_{k}}\otimes \bD_{1,k} & -\vect(\bD_{2,k}) \\ -(\vect(\bD_{2,k}))^{*} & 1\end{pmatrix}\bY\right)\nonumber\\
\mbox{s.t.} &&\tr(\bY) = p^{\max}_{\rT} + 1,\\
&& \tr\left(\begin{pmatrix}\bI_{d_{k}}\otimes\bD_{3,k} & \b0_{N_{\rT,k}d_{k} \times 1} \\ \b0_{1 \times N_{\rT,k}d_{k}} & 1\end{pmatrix}\bY\right) = \eta_{\rT,k} + 1,\nonumber\\\\
&& \tr\left(\begin{pmatrix}\b0_{N_{\rT,k}d_{k}\times N_{\rT,k}d_{k}} & \b0_{N_{\rT,k}d_{k}\times 1} \\ \b0_{1 \times N_{\rT,k}d_{k}} & 1\end{pmatrix} \bY\right) = 1,\\
&& \bY \succeq \b0,~\rank(\bY) = 1.
\end{eqnarray}
Note that $(\cWMSE\mbox{-}\rnoPC\mbox{-}\bff_{k}\bff^{*}_{k})$ has the same form as $(\cTL\mbox{-}\bu_{m}\bu_{m}^{*})$ but with one more constraint. Since $(\cWMSE\mbox{-}\rnoPC\mbox{-}\bff_{k}\bff^{*}_{k})$ has $n = 3$ constraints (excluding the rank-one constraint), its SDR obtained by relaxing the rank-one constraint is exact~\cite{HuangPalomar2010a:TSP}. Thus, we can use the same steps as those in solving $(\cTL\mbox{-}\bu_{m}\bu_{m}^{*})$ to find the desired column vector $\bff_{k}$ corresponding to a rank-one global optimum of $(\cWMSE\mbox{-}\rnoPC\mbox{-}\bff_{k}\bff^{*}_{k})$. We then use the $\vect^{-1}$ operator to get a globally optimal solution $\bF_{k}$ of $(\cWMSE\mbox{-}\rnoPC\mbox{-}\bF_{k})$ from the resulting $\bff_{k}$.

\subsection{Algorithm for Sum MSE Minimization with Power Control $(\cWMSE\mbox{-}\rPC)$} \label{subsec:SRPCalgorithm}
In this section, we discuss briefly how to solve $(\cWMSE\mbox{-}\rPC)$. Using the same steps as in Section \ref{subsec:SRnoPCalgorithm}, we can develop Algorithm 3 for finding high-quality solutions of $(\cWMSE\mbox{-}\rPC)$. The details of Algorithm 3 are provided in~\cite{TruongHeath2011:Asilomar}, thus here we only compare and contrast the steps of Algorithm 3 and those of Algorithm 2. First, the matrix weight and receive filter designs for $(\cWMSE\mbox{-}\rPC)$ are exactly the same as those for $(\cWMSE\mbox{-}\rnoPC)$. Second, the relay processing matrix design for $(\cWMSE\mbox{-}\rPC)$ can be solved by the Lagrangian multiplier method with the only difference is that the multiplier must be nonnegative. Finally, the optimization problem for the transmit precoder design for $(\cWMSE\mbox{-}\rPC)$ is obtained by replacing the equality constraints in $(\cWMSE\mbox{-noPC}\mbox{-}\bF_{k})$ by the corresponding inequality constraints. Fortunately, the resulting optimization problem is convex with respect to $\bF_{k}$. In particular, it follows from Remark \ref{remark:WMSEconvexity} that the objective function of the problem $(\cWMSE\mbox{-PC}\mbox{-}\cF_{k})$ is convex with respect to $\bF_{k}$. In addition, since $\bD_{3,k}$ is a Hermitian and positive semidefinite matrix, then we can easily check that the constraints of the resulting problem are also convex with respect to $\bF_{k}$. Thus, any available software package for convex optimization could be used to solve for its unique global optimum $\bF_{k}$.

\subsection{Discussion}\label{subsec:discussion}
In this section, we discuss the proposed algorithms in the following aspects: i) the convergence, ii) the quality of the solution, and iii) the assumption on power constraints at the relays.

In terms of convergence, we are able to find a global optimum of the single-variables minimization problem in each iteration. Thus, the cost function of the original multi-variable optimization problem is non increasing after each iteration~\cite{Bertsekas1999:BOOK,BezdekHathaway2003:NPSC}. This guarantees that all the proposed algorithms are convergent. Note that the authors of~\cite{ShiEtAl2011:TSP} adopt the alternating minimization to develop an iterative algorithm for solving a weighted sum-MSE minimization problem for the single-hop MIMO interference broadcast channel. That optimization problem has a differentiable objective function and a set of separable constraints in the main variables. Using the results from the general optimization~\cite{Solodov1998:SIAM}, the authors of~\cite{ShiEtAl2011:TSP} are able to claim that their proposed alternating minimization algorithm converges to a stationary point of the corresponding weighted sum-MSE minimization problem, which is also a stationary point of the associated sum-rate maximization problem Nevertheless, the optimization problems in the paper, $(\cTL)$, $(\cWMSE\mbox{-}\rnoPC)$, and $(\cWMSE\mbox{-}\rPC)$, have non-separable constraints due to the impact of transmitter precoders on the transmit power constraints at the relays. Thus, we are currently unable to make any strong claim about whether or not our proposed iterative algorithms always converge to a stationary point of the corresponding optimization problems.

All the proposed algorithms are not guaranteed to reach a global optimum of the corresponding multi-variable optimization problem. The quality of the resulting solution depends on the initialization. One way to improve the performance of the proposed algorithms is to use multiple initializations, selecting the one with the best performance at the expenses of running time.

The proposed algorithms in the current form are applicable only under the assumption of sum-power constraint at the relays. If individual power constraints at the relays are considered, we must formulate the corresponding optimization problems. For example, as in Section \ref{sec:problemFormulation}, to find high-quality solutions of the sum-rate maximization problem with individual power inequality constraints at the relays, we can formulate the following weighted sum-MSE minimization problem 
\begin{eqnarray}
(\cWMSE\mbox{-}\rPC\mbox{-ind}):~\min_{\{\bF\},\{\bU\},\{\bW\},\{\bV\}} && WMSE_{\rsum}\big(\{\bF\},\{\bU\},\{\bW\},\{\bV\}\big)\label{eq:WMMSE1ind}\nonumber\\
\mbox{s.t.} 
&& p_{\rT,k} \leq p^{\rmax}_{\rT,k}, k = 1, \cdots, K, \label{eq:WMMSE3ind}\\
&& p_{\rX,m} \leq p^{\rmax}_{\rX,m}, m = 1, \cdots, M.\label{eq:WMMSE4ind}
\end{eqnarray}
In some cases, we can use the same steps as in the previous sections to develop new alternative minimization based algorithms for solve the counterpart problems with individual power constraints at the relays, like $(\cWMSE\mbox{-}\rPC\mbox{-ind})$. Note that the main difference between the sum power constraint case and the individual power constraint case is the extra constraints in the transmit precoder design. Specifically, with per-relay power constraints, the number of quadratic constraints of the resulting QCQP for the transmit precoder design is $(M+1)$ instead of 2 as with the sum-power constraint. The extra constraints make it impossible 
to use the same SDP method to a globally optimal solution of the transmit precoder design problems for individual relay power constraints when there are more than two relays, i.e., $M \geq 3$. Developing new methods to solve the counterpart problems to $(\cTL)$ or $(\cWMSE\mbox{-noPC})$ when $M \geq 3$ is left for future work. When power control is considered, however, we can still use the same steps as in the previous sections to solve for high-quality solutions to $(\cWMSE\mbox{-}\rPC\mbox{-ind})$. Indeed, with power control, the single-variable optimization problems for designing the relay processing matrices and transmit precoders in solving $(\cWMSE\mbox{-}\rPC\mbox{-ind})$ are convex~\cite{TruongHeath2011:Asilomar}. Thus, we are always able to find their global optimum.

\section{Simulations}\label{sec:Simulations}
This section presents Monte Carlo simulation results to investigate the average end-to-end sum-rate performance and to gain insights into the achieved multiplexing gains of the proposed algorithms. We consider only symmetric systems, which are denoted as $(N_{\rR} \times N_{\rT}, d)^{K} + N_{\rX}^{M}$, where $N_{\rR, k} = N_{\rR}$, $N_{\rT, k} = N_{\rT}$, $d_{k} = d$ and $N_{\rX, m} = N_{\rX}$ for $k \in \cK, m \in \cM$. The power values are normalized such that $\sigma_{\rR, k} = \sigma_{\rX, m} = 1$,  $p^{\max}_{\rT,k} = P$, $p^{\max}_{\rX,m}= P$, and $p^{\max}_{\rX} = M P$ for $k \in \cK, m\in \cM$. The channel realizations are flat in time and frequency. The channel coefficients are generated as i.i.d. zero-mean unit-variance complex Gaussian random variables. No path loss is assumed in the simulations, thus the average power of all cross-links on the same hop is the same. The plots are produced by averaging over 1000 random channel realizations. For each channel realization, the initial transceivers are chosen randomly subject to the power constraints at the transmitters and relays. The same initializations are used where applicable. Each iteration updates either one transmitter or one relay and then all the receive filters. The same order of relays or transmitters selected for updating is used where applicable, for example, in the comparison of the proposed algorithms. We use the CVX toolbox~\cite{CVX:Software} to solve convex problems. 

For comparison, we consider the dedicated DF relay interference channel where one DF relay is dedicated to aid one and only one transmitter-receiver pair, i.e., $K = M$. Using equal time-sharing, the end-to-end achievable rate of a pair is defined as half of the minimum between the achievable rate from the transmitter to the associate DF relay and that from the relay to the receiver. We are interested only in the performance of DF relays when spatial interference alignment strategies, like those in~\cite{PetersHeath2008:ICASSP, ShiEtAl2011:TSP}, are applied on two hops. Although other interference alignment techniques, like asymmetric complex signaling~\cite{CadambeEtAl2010:TIT}, may improve the performance of DF relays, their impacts on DF relays are left for future work. Based on~\cite{RazaviyaynEtAl2011:TIT}, we derive an upper-bound on the achievable end-to-end multiplexing gain of the dedicated DF relay interference channel $(N_{\rR} \times N_{\rT}, d)^{K} + N_{\rX}^{K}$ as $ 0.5 * \min\left\{\left\lfloor \frac{K(N_{\rX} + N_{\rT})}{K+1}\right\rfloor, \left \lfloor \frac{K(N_{\rR} + N_{\rX})}{K+1}\right\rfloor\right\}$. Individual power constraints at the relays are considered in the comparison of AF relays and DF relays. We also assume the transmit power at a transmitter or a relay in DF relay systems is equal to $P$.

\subsubsection{Convergence}
\begin{figure}[h]
\centering
\subfigure[Total leakage power at the receivers over iterations of Algorithm 1 for a channel realization of $(4\times 4, 2)^{3}+ 4^{3}$.]{
\includegraphics[width=2.9in]{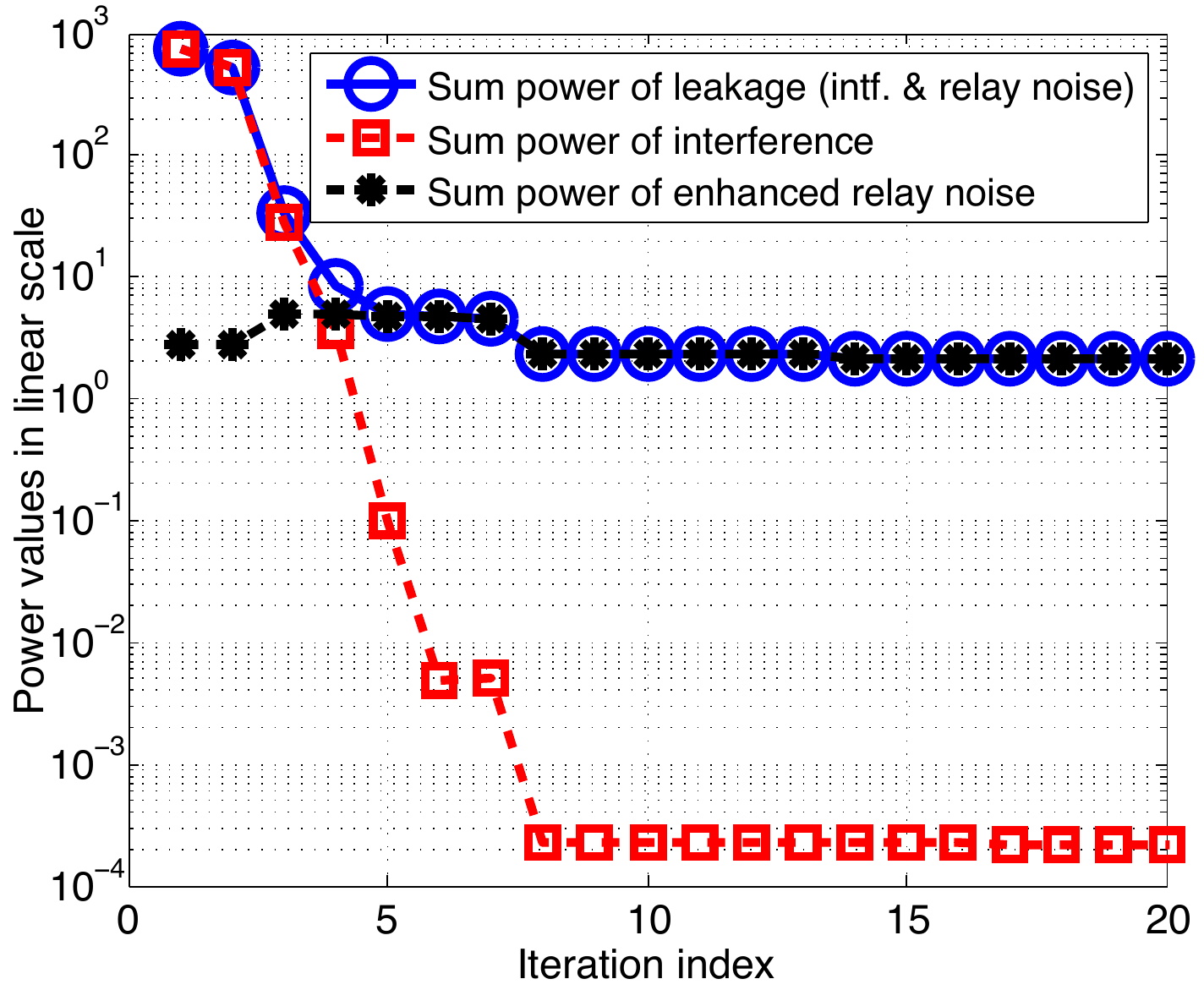}
\label{subfig:SumLeakageConvergence20B}
}
\hspace{5pt}
\subfigure[Matrix-weighted sum-MSE values over iterations of Algorithm 2 and Algorithm 3  for a channel realization of $(2\times 4, 1)^{4}+ 2^{4}$.]{
\includegraphics[width=2.9in]{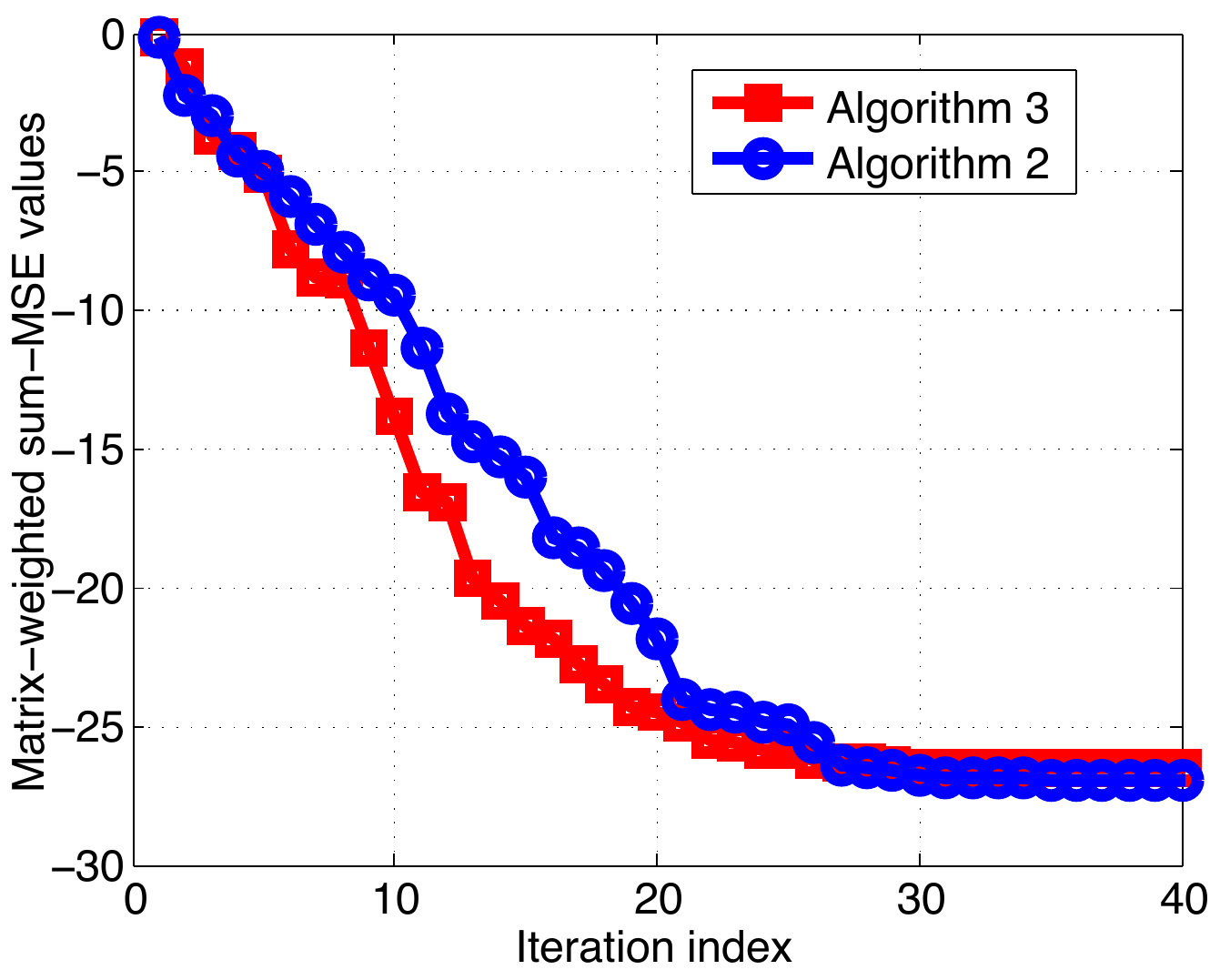}
\label{subfig:WMSEconvergence}
}
\vspace{-5pt}
\caption{Convergence behavior of the proposed algorithms.}
\label{fig:Convergence}
\vspace{-10pt}
\end{figure}

Fig. \ref{fig:Convergence} illustrates the convergence behavior of the proposed algorithms. Fig. \ref{subfig:SumLeakageConvergence20B} provides the analysis of the sum power of post-processed leakage signals of Algorithm 1 for a random channel realization of the $(4\times 4, 2)^{3}+ 4^{3}$ system. We observe that the sum power of leakage signals decreases monotonically over iterations. Interestingly, the interference and the enhanced relay noise change their roles during the process of Algorithm 1. The interference is dominant at the beginning, however, it can be aligned and then cancelled quickly in a few iterations. After this point, the enhanced relay noise becomes dominant - its sum power is thousands times larger than the interference sum power. Unfortunately, given that many spatial dimensions have been devoted to deal with interference, it becomes challenging for Algorithm 1 to align and cancel the enhanced relay noise power. Intuitively, the enhanced relay noise can be thought of as a source of single-hop interference from ``virtual uncoordinated relays'' that impacts directly the receivers. Thus, we need to take into account both the interference and enhanced relay noise in the design of interference alignment strategies for the AF relay interference channel. Fig. \ref{subfig:WMSEconvergence} provides the values of $WMSE_{\rsum}$ achieved by Algorithm 2 and by Algorithm 3 over iterations for a channel realization of the $(2 \times 4, 1)^{4} + 2^{4}$ system. We observe that $WMSE_{\rsum}$ values for both algorithms are non increasing over iterations. Although the convergence speeds of the proposed algorithms are quite fast for these configurations, they might be slow for networks with large values of $K$ or $d$.

\subsubsection{Comparison of the Proposed Algorithms}

\begin{figure}[h]
\centering
\includegraphics[width=2.9in]{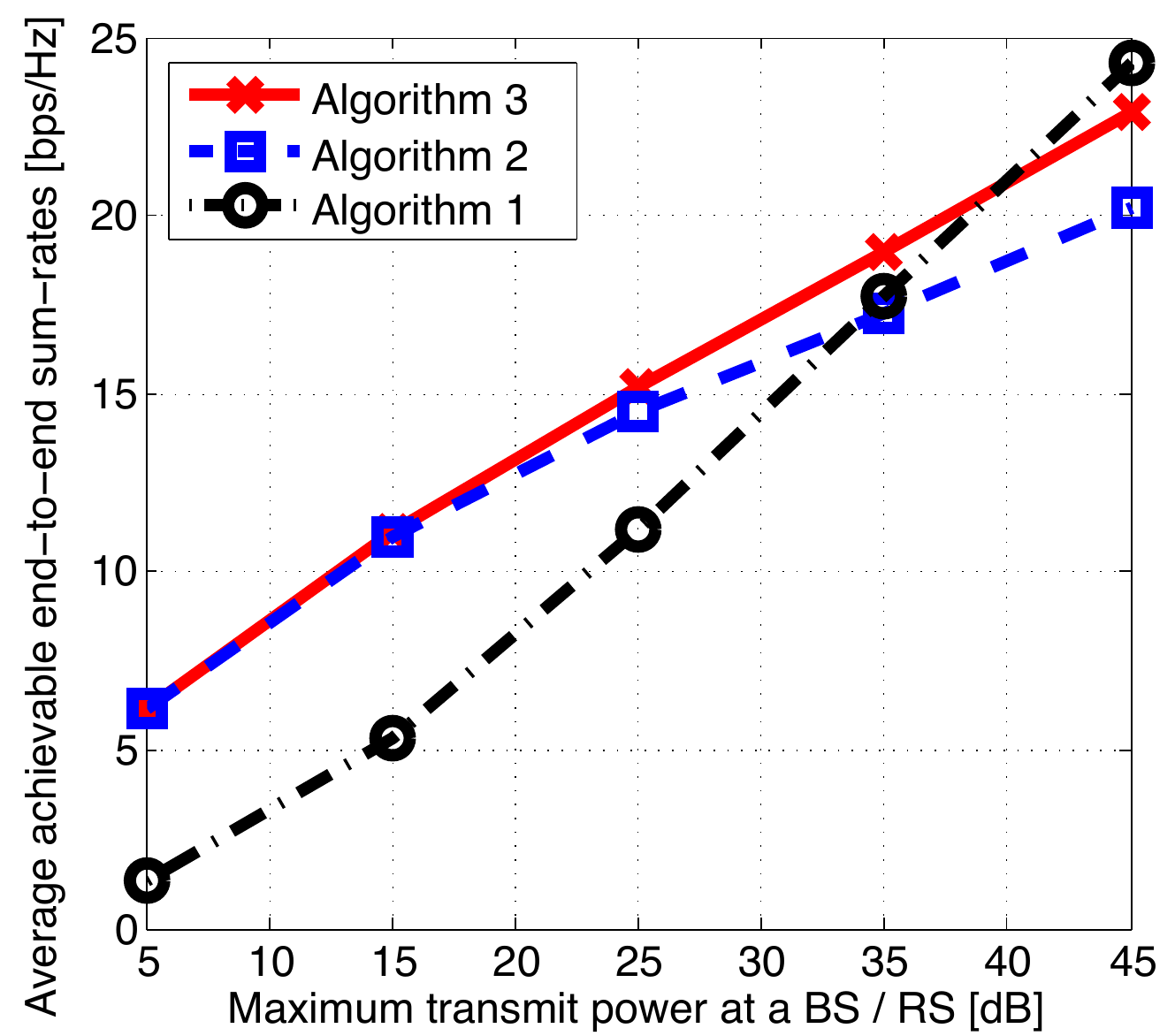}
\caption{Comparison of the average achievable end-to-end sum-rates of the proposed algorithms for the $(2\times 4, 1)^{4}+ 2^{4}$ system.}
\label{fig:algComparison1}
\vspace{-10pt}
\end{figure}

In this experiment, we simulate the average achievable end-to-end sum-rates for the $(2\times 4, 1)^{4}+ 2^{4}$ system under the sum power constraint at the relays as shown in Fig. \ref{fig:algComparison1}. We consider a sum relay power constraint. Thanks to power control, Algorithm 3 outperforms Algorithm 2 in this experiment. Both Algorithm 2 and Algorithm 3 outperform Algorithm 1 at low-to-medium SNR values because Algorithm 1 does not take into account the desired signal and noise at the receivers while the other do. Interestingly, at high SNR values, Algorithm 1 outperforms both Algorithm 2 and Algorithm 3. Especially, Algorithm 1 can achieve a higher multiplexing gain than do the other. Zooming in on per-user achievable end-to-end rates, we find that for Algorithm 2 and Algorithm 3, some users have much smaller rates than do the others; they even turn off some data streams. This unfairness limits the maximum end-to-end multiplexing gains achievable by the two algorithms. Thus, Algorithm 1 is more suitable than the others for investigating the maximum achievable end-to-end multiplexing gains of MIMO AF relay networks.

\subsubsection{Sum Power Constraints vs. Individual Power Constraints at Relays} 
\begin{figure}[h]
\centering
\includegraphics[width=2.9in]{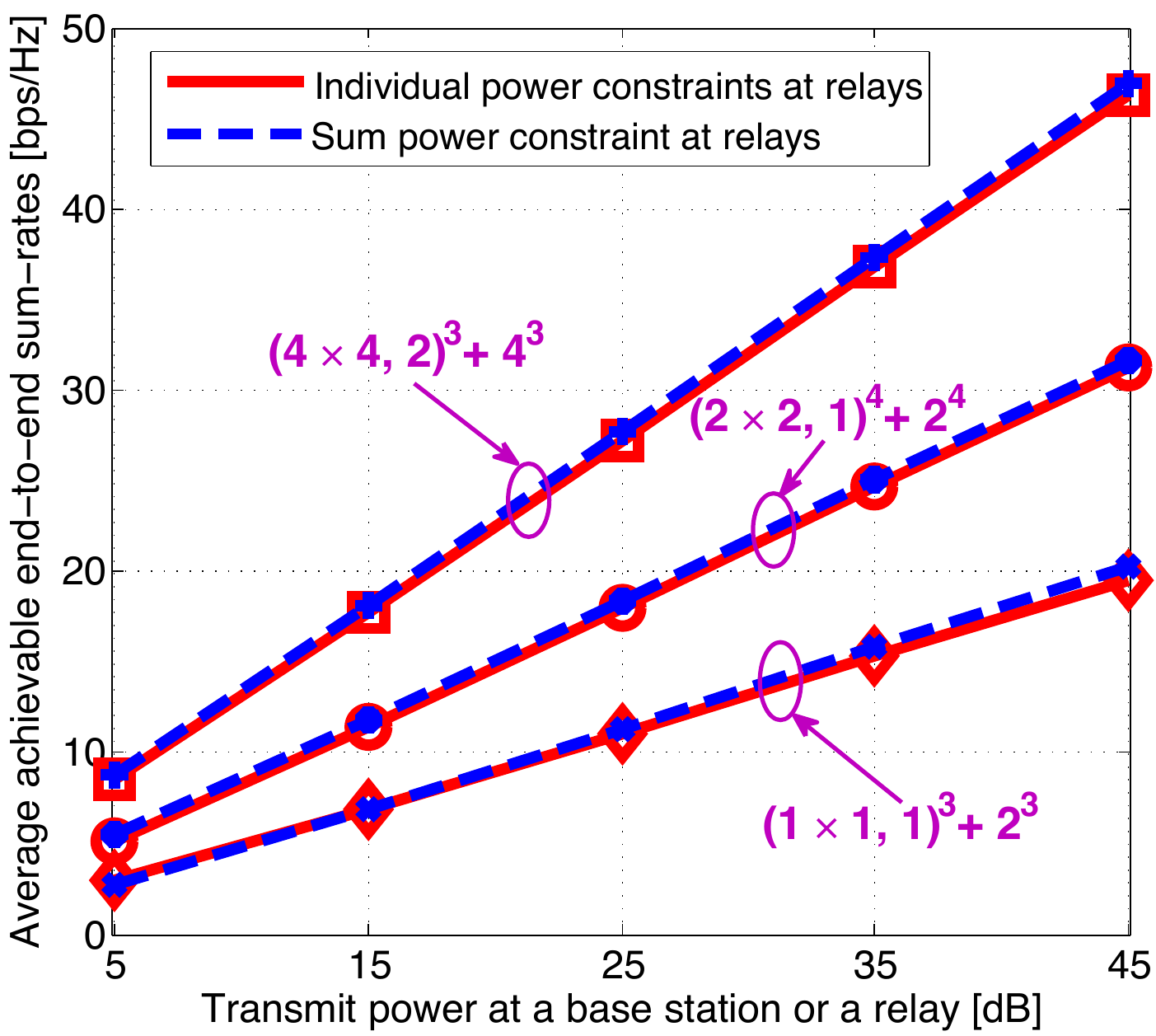}
\caption{Achievable end-to-end sum-rates with both types of power constraints at the relays for the $(1\times 1, 1)^{3}+ 2^{3}$, $(2\times 2, 1)^{4}+ 2^{4}$, and $(4\times 4, 2)^{3}+ 4^{3}$ systems.}
\label{fig:powerConstraintImpact}
\vspace{-10pt}
\end{figure}

In the previous experiments, we consider sum-power constraints at the relays. In this experiment, we consider the impacts of individual power constraints. Note that any feasible point satisfies the individual power constraints at the relays also satisfies the corresponding sum-power constraint. Based on the discussion in Section \ref{subsec:discussion}, we focus on the case where power control is considered as it allows for the use of any number of relays. Fig. \ref{fig:powerConstraintImpact} shows the achievable end-to-end sum-rates as functions of the transmit power at a base station or a relay for both types of power constraints at the relays for the following three systems: i) $(4\times 4, 2)^{3}+ 4^{3}$, ii) $(2\times 2, 1)^{4}+ 2^{4}$, and iii) $(1\times 1, 1)^{3}+ 2^{3}$. We observe that Algorithm 3 for the sum-power constraint case slightly outperforms its counterpart algorithm for the individual power constraint case in terms of maximizing average achievable end-to-end sum-rates. This gain is due to having more freedom in power allocation in the sum-power constraint case as relays may transmit at a higher value than the maximum transmit power at a relay in the individual power constraint case. This means that extra constraints added by the individual power constraints at the relays have little impact on the end-to-end sum-rate performance of the proposed algorithms. In the following experiments, we use only the counterpart version of Algorithm 3 that is designed for individual relay power constraints, which we refer to as `modified Algorithm 3'.

\subsubsection{Comparison with Existing Strategies} In these experiments, we simulate several existing transceiver design strategies for the relay interference channel. For fair comparison, in this experiment and the remaining experiments, we consider the individual relay power constraints. Specifically, we simulate two strategies for the AF relay case. One is the AF TDMA distributed beamforming (BF), where all the relays help only one transmitter-receiver pair at a time (which is an extension of the design in~\cite{LiangSchober2010:VTC} for multiple-antenna receivers). Another is the dedicated relay BF where each AF relay is devoted to aiding one and only one transmitter-receiver pair. This means that interference is ignored and we apply the joint source-relay design in~\cite{FangEtAl2006:SAM, RongEtAl2009:TSP} independently for the two-hop channels from the transmitters to their associated receivers. We also three strategies for the DF relay case that correspond to independent applications of single-hop strategies on two hops. The single-hop strategies include the following: i) selfish (SF) beamforming (i.e., each transmitter aims at maximizing the achievable rate to its associated receiver), ii) interference alignment strategy based on total leakage (TL) minimization~\cite{PetersHeath2008:ICASSP}, and iii) the iteratively weighted MSE sum-rate (SR) maximization strategy~\cite{ShiEtAl2011:TSP}.

\begin{figure}[h]
\centering
\includegraphics[width=2.9in]{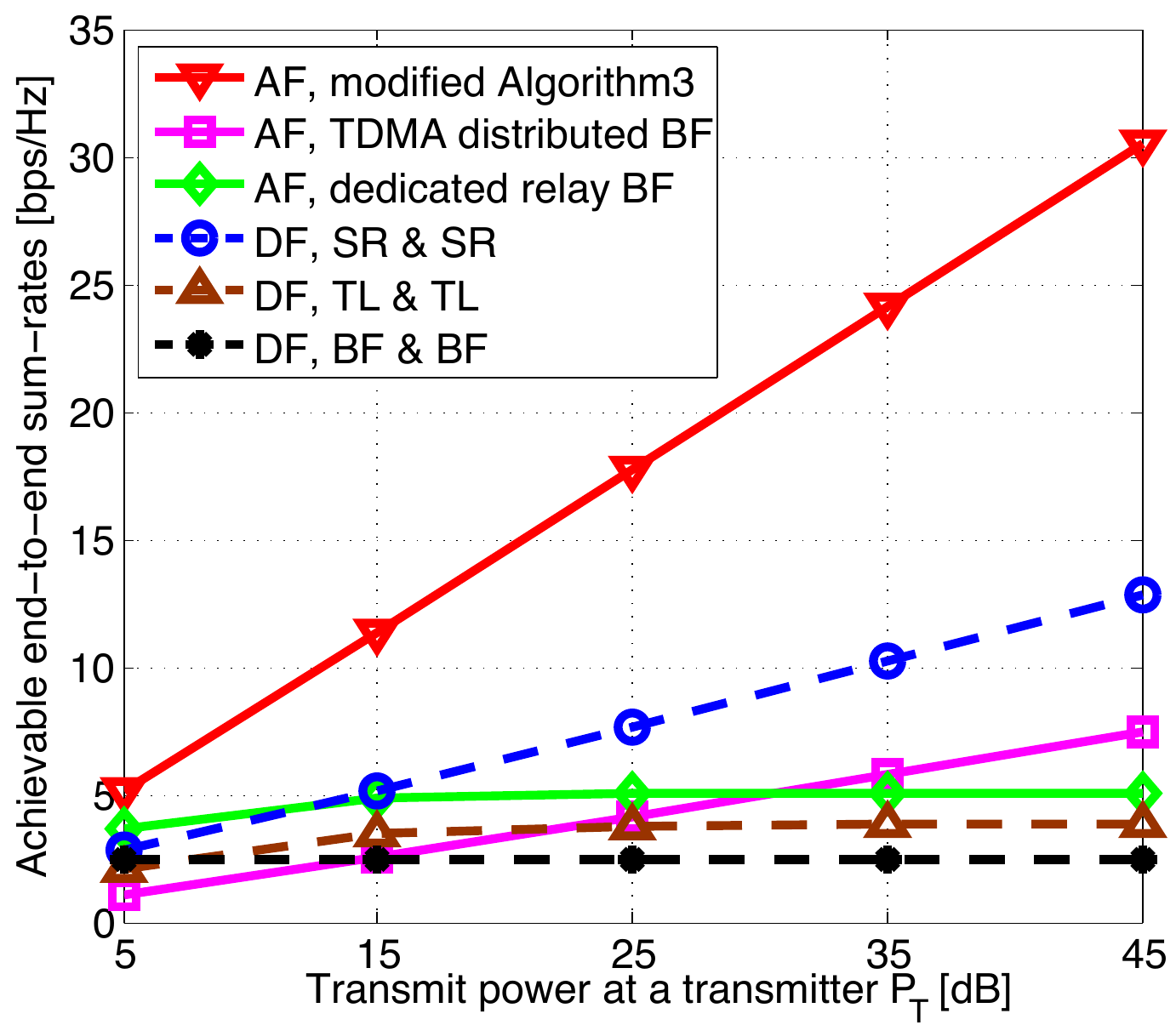}
\caption{Achievable end-to-end sum-rates for the $(2\times 2, 1)^{4}+ 2^{4}$ system.}
\label{fig:IndComparison02020204040401Simplified}
\vspace{-10pt}
\end{figure}

Fig. \ref{fig:IndComparison02020204040401Simplified} shows the results for the $(2\times 2, 1)^{4}+ 2^{4}$ system. Recall that Algorithm 3 outperforms all the other in all regions. It achieves an end-to-end multiplexing gain of 2 (which is equal to half of the total number of data streams). Note that we do not claim that this is the maximum degrees of freedom of this system. More complicated designs, for example those that can take advantage of symbol extensions~\cite{GouEtAl2010:Arxiv}, may achieve higher end-to-end multiplexing gains. Unaware of interference, the dedicated relay strategies for both AF relay and DF relay cases achieve zero multiplexing gains. While the multiplexing gain achieved by the DF TL $\&$ TL strategy is zero, that by the DF SR $\&$ SR strategy is nonzero. The reason is that interference alignment is not feasible for the configuration on the two hops, interference cannot be completely eliminated using the TL algorithm. Although the SR algorithm is able to turn off some data streams, one data stream on each hop in this case, to make interference alignment feasible. Note, however, that it may turn off data streams of different pairs on two hops. Thus, on average the DF SR $\&$ SR strategy achieves an end-to-end multiplexing gain less than 1.5 (half of the number of remaining data streams when interference alignment is feasible). Finally, thanks to orthogonalization transmission, the AF TDMA distributed BF can achieve an end-to-end multiplexing gain of 0.5.

\begin{figure}[h]
\centering
\vspace{-5pt}
\includegraphics[width=2.9in]{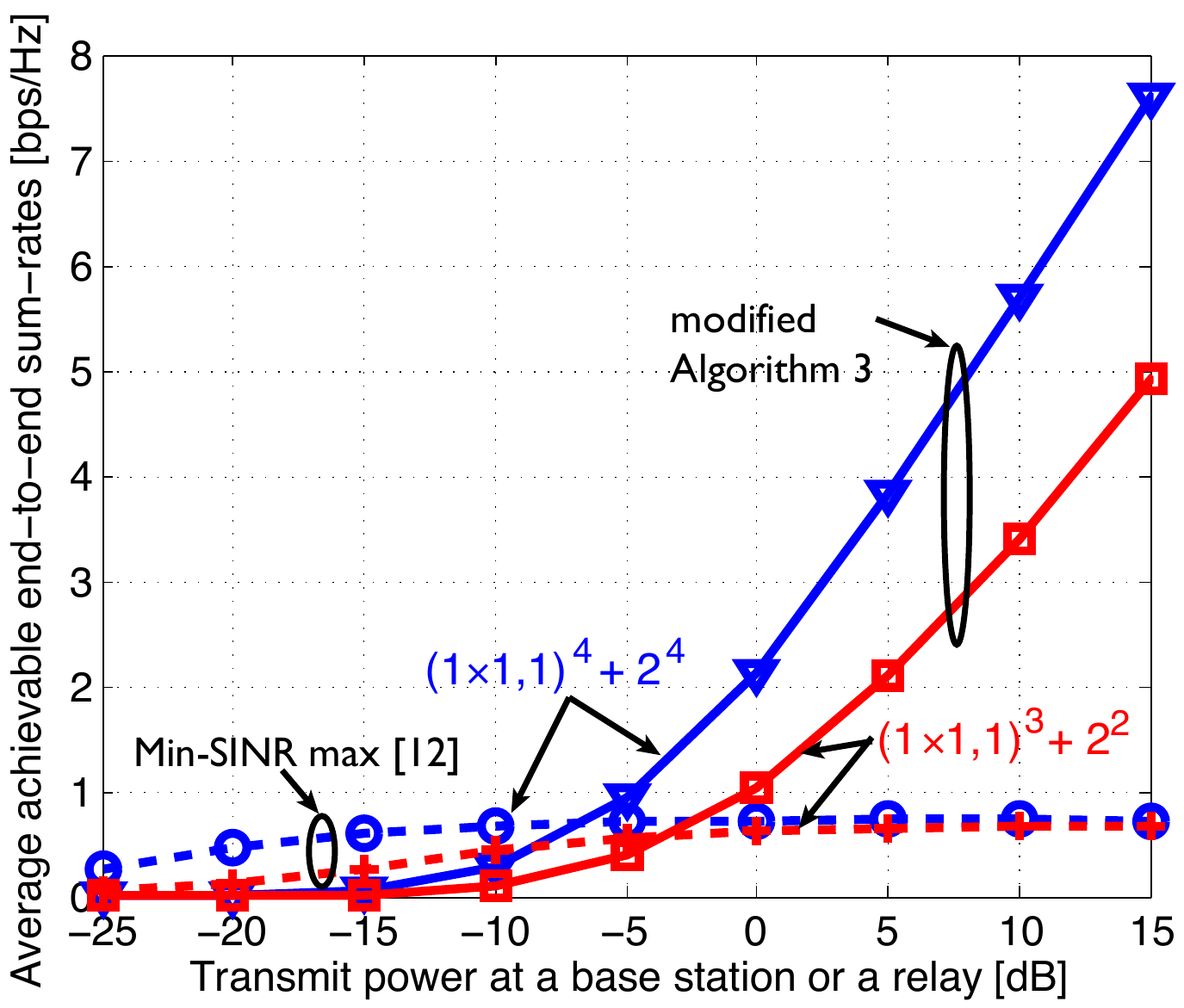}
\caption{Comparison of achievable end-to-end sum-rates of Algorithm 3 (shown by the solid lines) and the minimum-SINR maximization algorithm in~\cite{ChaliseVandendorpe2010:TSP} (shown by the dashed lines) for two configurations $(1\times 1, 1)^{4}+ 2^{4}$ and $(1\times 1, 1)^{3}+ 2^{2}$.}
\label{fig:AlgComparisonChalise}
\vspace{-10pt}
\end{figure}

Another experiment focuses on comparing Algorithm 3 with the the minimum-SINR maximization algorithm in~\cite{ChaliseVandendorpe2010:TSP}. Note that the algorithm in~\cite{ChaliseVandendorpe2010:TSP} is applicable only for single-antenna receivers. Fig. \ref{fig:AlgComparisonChalise} shows the end-to-end achievable sum-rates of Algorithm 3 and the minimum-SINR maximization algorithm for two configurations $(1\times 1, 1)^{4}+ 2^{4}$ and $(1\times 1, 1)^{3}+ 2^{2}$. We observe that the end-to-end sum-rate performance of the minimum-SINR maximization algorithm increases at low transmit power (i.e., in the noise-limited regime) and saturates at high transmit power (i.e., the interference-limited regime). Thus, the algorithm achieves a end-to-end multiplexing gain of zero. This is reasonable since it is not designed specifically for interference management. Thanks to its capability of interference management, Algorithm 3 still achieves non-zero end-to-end multiplexing gains and provides large end-to-end sum-rate gains over the minimum-SINR maximization algorithm in the interference-limited regime. Our algorithm, however, performs worse than the the minimum-SINR maximization algorithm in the noise-limited regime where interference becomes a negligible issue.

\subsubsection{Maximum Achievable Multiplexing Gains}
\begin{figure}[h]
\centering
\includegraphics[width=2.9in]{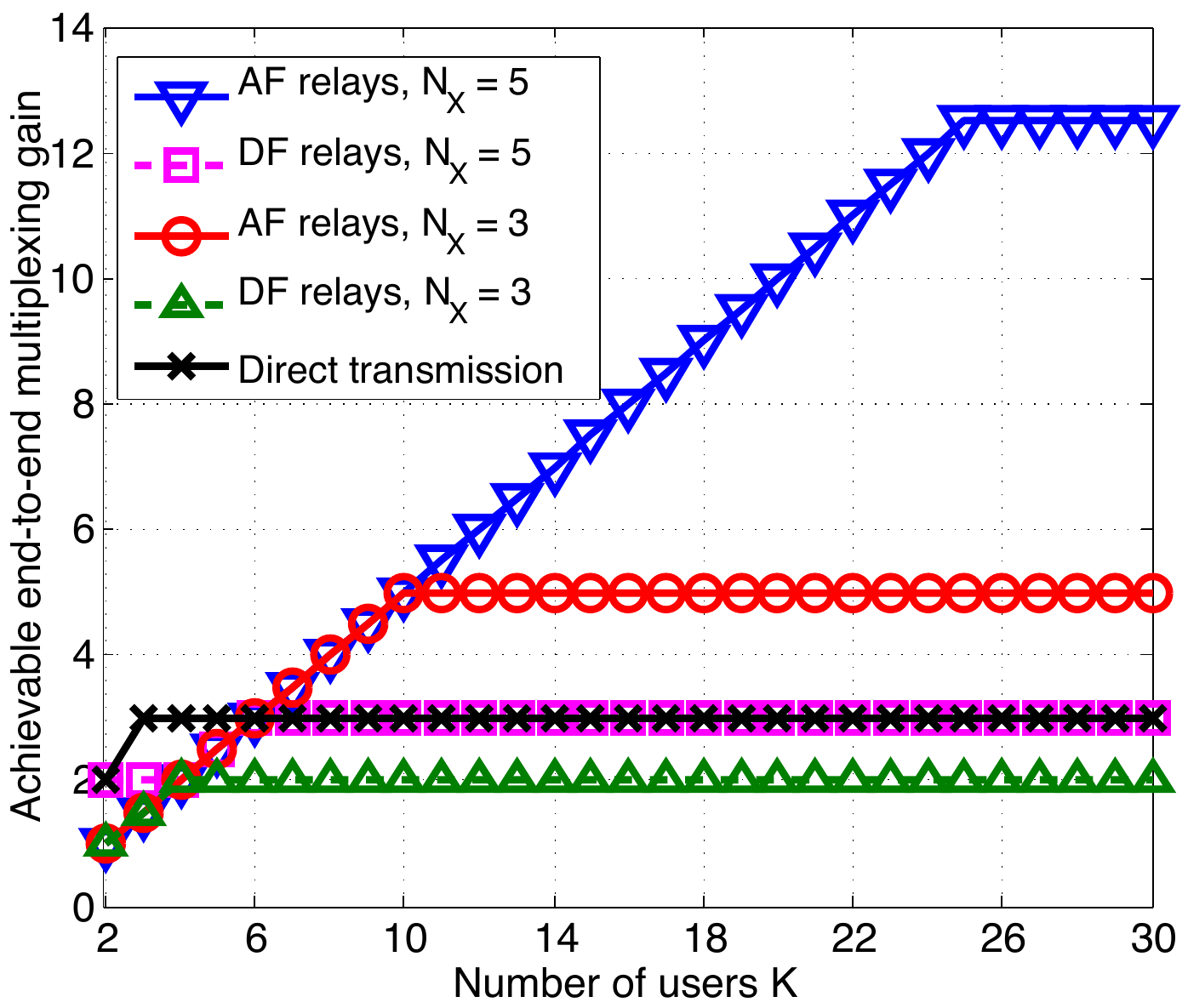}
\caption{Achievable end-to-end multiplexing gains as functions of $K$ for the $(2\times 2, 1)^{K}+ N_{\rX}^{K}$ systems.}
\label{fig:DoFvsK}
\vspace{-15pt}
\end{figure}

We fix $N_{\rR} = N_{\rT} = 2$ and $d = 1$. Fig. \ref{fig:DoFvsK} shows the achievable end-to-end multiplexing gains achieved by using Algorithm 1 as a function of $K$ for $N_{\rX} = 3$ and $N_{\rX} = 5$ for AF relays, DF relays, and direct transmission. We notice that with these values of $N_{\rX}$ and when $K$ is small, due to the half-duplex loss, both the AF relay and DF relay cases achieve lower multiplexing gains than the direct transmission. While the DF relay case cannot outperform the direct transmission, the AF relay case can achieve higher multiplexing gains when there are more than 6 users. Thus, we can claim that AF relays help increase the achievable end-to-end multiplexing gains of interference channels. In addition, we observe that there exist upper-bounds on the achievable end-to-end multiplexing gains for all the simulated cases - AF relays, DF relays, and direct transmission. Theoretical investigation of the upper-bounds is left for future work.

\subsubsection{Opportunistic Approach}
\begin{figure}[t]
\centering
\includegraphics[width=2.9in]{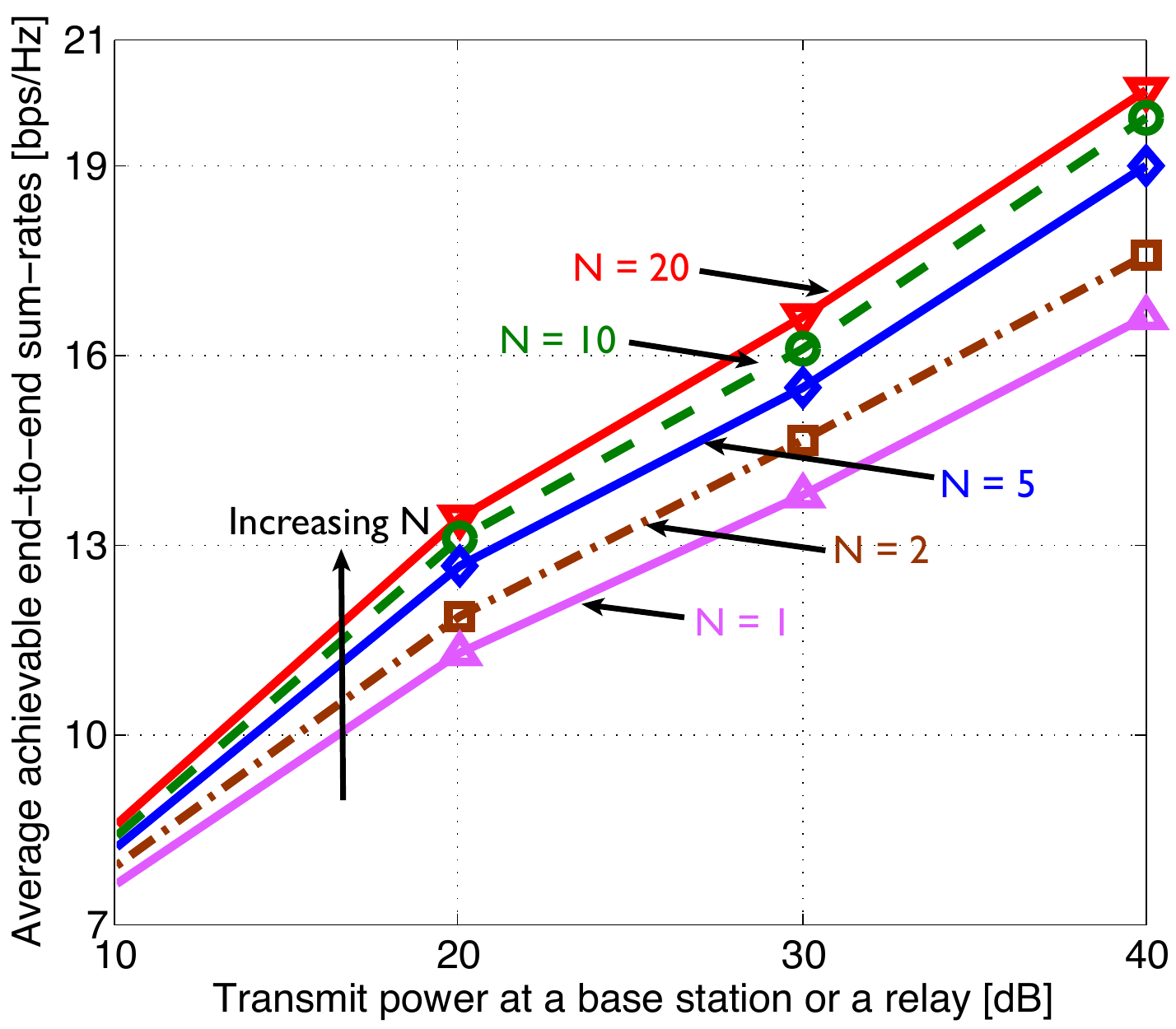}
\caption{Achievable end-to-end sum-rates of the opportunistic approach for the $(2\times 2, 1)^{4}+ 2^{4}$ system with $N = \{1, 2, 5, 10, 20\}$.}
\vspace{-15pt}
\label{fig:opportunistic}
\end{figure}

The end-to-end sum-rate performance of the stationary points found by Algorithm 2 and Algorithm 3 depend significantly on the initializations. The opportunistic approach proposes to use multiple initializations and then chooses the one with the highest end-to-end sum-rates. Let $N$ denote the number of random initializations. Fig. \ref{fig:opportunistic} shows the average end-to-end sum-rates for several values of $N$ achieved by Algorithm 3 in the $(2\times 2, 1)^{4}+ 2^{4}$ system.  For this setting, at a transmit power of 30dB, the gain provided by the opportunistic approach over the non-opportunistic approach is 6.4$\%$ for $N =2$, 13.2$\%$ for $N = 5$, 16.9$\%$ for $N=10$, and 20.6$\%$ for $N = 20$. Note that the higher the value of $N$, the larger the average achievable end-to-end sum-rates. Also, the additional gains obtained by using an extra random initialization decreases in $N$. Nevertheless, the benefits of this opportunistic approach come at the expense of longer running time.

\section{Conclusions and Future Work}\label{sec:Conclusions}
We developed three cooperative algorithms for joint designs of the transmitters, relays, and  receivers of the MIMO AF relay interference channel. Algorithm 1 aims at minimizing the sum power of the interference signals and the enhanced noise from the relays. Based on a relationship between MSE and mutual information, Algorithm 2 (Algorithm 3) is able to find a stationary point of the end-to-end sum-rate maximization problems with equality (inequality) power constraints. Simulations show that thanks to the consideration of the desired signal power and the noise power at the receivers, Algorithm 2 and Algorithm 3 outperform Algorithm 1 at low-to-medium SNR. Nevertheless, they perform worse than Algorithm 1 at high SNR due to unfairness in rate allocation among users. The multiplexing gains achievable by the proposed algorithms provide lower bounds on the total number of degrees of freedom in MIMO AF relay networks, which remains unknown. Also, the use of AF relays results in higher end-to-end multiplexing gains than both the use of DF relays and the direct transmission.

A major limitation of our algorithms is that global CSI is required to implement them in their present form. Naturally this is challenging to achieve in a distributed system. We believe the results are still valuable, however, because they provide a benchmark for developing algorithms that relax the global CSI assumptions. Future work should focus on developing cooperative algorithms that require less overhead, have faster convergence speed, allow for lower implementation complexity, and account for channel estimation error

\bibliographystyle{IEEEtran}
\bibliography{IEEEabrv,relayIA}

\end{document}